\documentclass[preprint,aip,jmp,showpacs,nofootinbib]{revtex4-1}		 

\usepackage{amsthm,amssymb,amsmath}				
\usepackage{graphicx,xcolor}					
\usepackage[colorlinks = true,linkcolor = blue, 
            urlcolor  = blue, citecolor = blue, 
            anchorcolor = blue]{hyperref}       

\begin{document}

\title{
$\kappa$-Deformed quantum and classical mechanics for a
system with position-dependent effective mass
}

\author{Bruno G. da Costa}
\email{bruno.costa@ifsertao-pe.edu.br}
\affiliation{Instituto Federal de Educa\c{c}\~ao, Ci\^encia e Tecnologia do
                 Sert\~ao Pernambucano, 56316-686 Petrolina-PE, Brazil}
\author{Ignacio S. Gomez}
\email{nachosky@fisica.unlp.edu.ar}
\affiliation{Instituto de F\'{i}sica, Universidade Federal da Bahia,
			 Rua Barao de Jeremoabo, 40170-115 Salvador-BA, Brazil}
\author{Mariela Portesi}
\email{portesi@fisica.unlp.edu.ar}
\affiliation{IFLP, CONICET, UNLP, Diagonal 113 e/63 y 64, 1900 La Plata, Argentina}
\affiliation{Facultad de Ciencias Exactas, Universidad
	Nacional de La Plata, C.C. 67, 1900 La Plata, Argentina}

\date{\today}

\begin{abstract}
We present the quantum and classical mechanics formalisms for a particle with
position-dependent mass in the context of a deformed algebraic structure
(named $\kappa$-algebra), motivated by the Kappa-statistics.
From this structure we obtain deformed versions of the position and momentum operators,
which allow to define a point canonical transformation that maps a particle with
constant mass in a deformed space into a particle with  position-dependent mass
in  the standard space. We illustrate the formalism with a particle confined
in an infinite potential well and the Mathews-Lakshmanan oscillator,
exhibiting uncertainty relations depending on the deformation.
\end{abstract}


\maketitle

\section{Introduction}

Minimum length scales are of crucial importance in several areas of
physics like quantum gravity, string theory, relativity, fundamentally
due to the techniques developed for removing divergences
in field theories maintaining the parameters lengths as
universal constants of the theory in question
(for a review see for instance
Ref.~\onlinecite{Hos-2013}).
In this sense, the seek for these minimum lengths in
quantum mechanics has been translated into
generalizations of the standard commutation relationship
between position and momentum.
\cite{Kem-1994}
Further studies in noncommuting quantum spaces led to
a Schr\"odinger equation with a
position-dependent effective
mass (PDM).
\cite{Costa-Filho-2016}
Along the last decades the PDM systems
have attracted attention because of
their wide range of applicability in
semiconductor theory,
\cite{vonroos_1983,
Lima-Vieira-Furtado-Moraes-Filgueiras-2012,
Glasser-2020,Costa-Gomez-Santos-2020}
nonlinear optics,
\cite{Khordad}
quantum liquids,
\cite{Saavedra_1994,Barranco-1997}
inversion potential for NH$_3$
in density functional theory,
\cite{Aquino_1998}
particle physics,
\cite{Bethe_1986}
many body theory,
\cite{Bencheikh-et-al-2004}
molecular physics,
\cite{Christiansen-Cunha-2014}
Wigner functions,
\cite{Cherroud-2017}
relativistic quantum mechanics,
\cite{Aydogdu-Arda-Sever-2012}
superintegrable systems,
\cite{Ranada-2016}
nuclear physics,
\cite{Alimohammadi-Hassanabadi-Zare-2017}
magnetic monopoles,
\cite{Schmidt-2018,Jesus-2019}
astrophysics,
\cite{Richstone_1982}
nonlinear oscillations,
\cite{Mathews-Lakshmanan-1974,
Mathews-Lakshmanan-1975,
Tiwari-2013,
Bagchi-2015,
Carinena-2015,
Amir-Iqbal-2015,
Ruby-2015,
Schulze-Halberg-Roy-2016,
Karthiga-et-al-2017,
Khlevniuk-2018}
factorization methods and supersymmetry,
\cite{Bravo-PRD-2016,Mustafa-2020,Midya-Roy-2009,
Amir-Iqbal-2016,Karthiga-2018}
coherent states,
\cite{Amir-Iqbal-2016-CS,Tchoffo-2019,Ruby-Senthilvelan-2010}
etc.

Complementarily,
it has been found that the mathematical foundations of the PDM systems
rely on the assumption of the noncommutativity
between the mass operator $m(\hat{x})$ and the linear momentum operator $\hat{p}$,
thus giving place to the ordering problem for
the kinetic energy operator,
\cite{vonroos_1983,Levy-Leblond-1995,BenDaniel-Duke-1966,
Gora-Williams-1969,Zhu-Kroemer-1983,Morrow-Brownstein-1984,Li-Kuhn-1993,
SouzaDutra-Almeida,Mustafa-Mazharimousavi-2007}.
In addition, the development of generalized translation
operators motivated the introduction of
a position-dependent linear momentum
for characterizing a particle with a PDM
\cite{Costa-Gomez-Santos-2020,
CostaFilho-Almeida-Farias-AndradeJr-2011,
Mazharimousavi-2012,
CostaFilho-Alencar-Skagerstam-AndradeJr-2013,
RegoMonteiro-Nobre-2013-PRA,
Costa-Borges-2014,
Arda-Server,
Braga-CostaFilho-2016,
Costa-Borges-2018,
Costa-Gomez-2020}
that can be related to a generalized algebraic structure
(called $q$-algebra
\cite{Borges_2004})
inherited from the mathematical background of nonextensive statistics.
\cite{Tsallis-Springer-2009}
Concerning these formal structures,
the $\kappa$-deformed statistics,
originated from the $\kappa$-exponential
and $\kappa$-logarithm functions,
allows to develop an algebraic structure,
called
$\kappa$-algebra,
\cite{kaniadakis-physa-2001,
kaniadakis-pre-2002,
kaniadakis-pre-2005,
Scarfone-2016-Legendre-transform,
kaniadakis-EPJA-2009,
Abe-2004,
Kaniadakis-Scarfone-Sparavigna-Wada-2017,
Scarfone-2015,
kaniadakis-Entropy-2013,
Scarfone-2017,
Wada-law-of-error-2006,
Wada-kappa_Gamma-2013,
Souza-2014,
Silva-Silva-Ramos-2020,
Scarfone-2018,
Kaniadakis-Lissia-Scarfone-2005}
with similar properties to the those of the $q$-algebra.
In particular, the $\kappa$-statistics has been employed in
plasma physics,
\cite{Lapenta-2007}
astrophysics,
\cite{Carvalho-2010}
paramagnetic systems,
\cite{Livadiotis-2016}
nonlinear diffusion,
\cite{Wada-2010}
social systems,
\cite{Clementi-2008}
complex networks,
\cite{Stella-2014}
analysis of human DNA,
\cite{Costa-Silva-Anselmo-Silva-2019}
blackbody radiation,
\cite{Ourabah-2014}
quantum entanglement,
\cite{Ourabah-2015}
etc.

In this work we employ the
$\kappa$-algebra for generalizing classical and quantum mechanics
with the aim of studying the properties of the resulting
noncommuting space originated by the deformation.
Between these properties we found that the $\kappa$-deformed space,
classical and quantum, allows to characterize a PDM system with
the mass being univocally determined by the $\kappa$-algebra.
The work is organized as follows.
In Section \ref{sec:kappa-algebra}, we review the properties of
the $\kappa$-algebra that are used in the forthcoming sections.
Next, we present in Section \ref{sec:kappa-deformed-dynamics}
the dynamics resulting from a generic PDM
and then we specialize with the mass function $m(x)$
associated to the $\kappa$-algebra. Here we obtain
the Schr\"odinger equation associated to the $\kappa$-derivative and
we show that all the standard properties remain to be valid in
the deformed structure such as the continuity equation,
the wave-function
normalization, the classical limit, etc.
In Section \ref{sec:particle-infinite-well}
we illustrate our proposal with a particle
in an infinite potential well.
In Section \ref{sec:kappa-deformed-oscillator},
we use the $\kappa$-deformed formalism to revisit
the problem of the Mathews-Lakshmanan oscillator.
\cite{Mathews-Lakshmanan-1974,
Mathews-Lakshmanan-1975,
Tiwari-2013,
Bagchi-2015,
Carinena-2015,
Amir-Iqbal-2015,
Ruby-2015,
Schulze-Halberg-Roy-2016,
Karthiga-et-al-2017,
Khlevniuk-2018}
Finally, in Section \ref{sec:conclusions}
we draw some conclusions
and outline future perspectives.

\section{\label{sec:kappa-algebra}
		 Review of the $\kappa$-algebra}

The $\kappa$-statistics emerges from a generalization of the Boltzmann-Gibbs
entropy derived by means of a kinetic interaction principle,
that allows to characterize nonlinear kinetics in particle systems
(see, for instance, Ref.~\onlinecite{kaniadakis-physa-2001} for more details).
In the last two decades several theoretical developments 
have shown that the $\kappa$-formalism preserve features as
Legendre transform in thermodynamics,
\cite{Scarfone-2016-Legendre-transform}
H-theorem,
\cite{kaniadakis-EPJA-2009}
Lesche stability,
\cite{Abe-2004}
composition law of the $\kappa$-entropy,
\cite{Kaniadakis-Scarfone-Sparavigna-Wada-2017}
among others.
The mathematical background of the $\kappa$-deformed formalism is
based on generalizations of the standard exponential
and logarithm functions, from which it is possible to introduce
deformed versions of algebraic operators and calculus,
\cite{kaniadakis-physa-2001,kaniadakis-pre-2002,kaniadakis-pre-2005}
trigonometric and hyperbolic functions,
\cite{Scarfone-2015,kaniadakis-Entropy-2013}
Fourier transform,
\cite{Scarfone-2017}
Gaussian law of error,
\cite{Wada-law-of-error-2006}
Stirling approximation and Gamma function,
\cite{Wada-kappa_Gamma-2013}
Cantor set,
\cite{Souza-2014}
Lambert $W$ function
\cite{Silva-Silva-Ramos-2020},
information geometry,
\cite{Scarfone-2018}
and other possible exponential and logarithm functions,
\cite{Kaniadakis-Lissia-Scarfone-2005}
etc.

More specifically,
the so-called $\kappa$-exponential is a deformation of
the ordinary exponential function,
defined by
\cite{kaniadakis-physa-2001,
kaniadakis-pre-2002,
kaniadakis-pre-2005}
\begin{equation}
\label{eq:kappa-exponential}
\exp_\kappa{u} \equiv \left( \kappa u + \sqrt{1+\kappa^2 u^2} \right)^{1/\kappa}
			   = \exp \left( \frac{1}{\kappa} \textrm{arcsinh} (\kappa u) \right),
\quad (\kappa \in \mathbb{R}).
\end{equation}
The inverse function of the $\kappa$-exponential
is the $\kappa$-logarithm, given by
\begin{equation}
\label{eq:kappa-logatithm}
 \displaystyle \ln_\kappa {u} \equiv \frac{u^\kappa - u^{-\kappa}}{2\kappa}
							   = \frac{1}{\kappa} \sinh (\kappa \ln u),
								 \quad (u > 0).
\end{equation}
In the limit $\kappa \rightarrow 0$,
the ordinary exponential and logarithmic functions
are recovered, {\it i.e.} $\exp_0 x = \exp x$ and $\ln_0 x = \ln x$.
These functions satisfy the properties
$\exp_\kappa (a) \exp_\kappa (b) = \exp_\kappa (a \overset{\kappa}{\oplus} b)$,
$ \exp_\kappa (a) / \exp_\kappa (b) = \exp_\kappa (a \overset{\kappa}{\ominus} b)$,
$ \ln_\kappa (a b) =  \ln_\kappa (a) \overset{\kappa}{\oplus} \ln_\kappa (b)$
and
$ \ln_\kappa (a/b) =  \ln_\kappa (a) \overset{\kappa}{\ominus} \ln_\kappa (b)$,
where the symbol $\overset{\kappa}{\oplus}$ represents the $\kappa$-addition operator
defined by
${a \overset{\kappa}{\oplus} b} \equiv a \sqrt{1+\kappa^2 b^2} +b \sqrt{1+\kappa^2 a^2}$,
and $\overset{\kappa}{\ominus}$ represents the $\kappa$-subtraction,
${a \overset{\kappa}{\ominus} b} \equiv a \sqrt{1+\kappa^2 b^2} -b \sqrt{1+\kappa^2 a^2}$.
\cite{kaniadakis-physa-2001,Scarfone-2015}

A $\kappa$-deformed calculus has been introduced in
Ref.~\onlinecite{kaniadakis-physa-2001} from the
deformed differential
\begin{equation}
d_\kappa u \equiv \displaystyle \lim_{u'\to u} u' \overset{\kappa}{\ominus} u
      = \frac{du}{\sqrt{1 + \kappa^2 u^2}} + \mathcal{O}((du)^2).
\end{equation}
The definition of a deformed variable $u_\kappa$
(also named deformed $\kappa$-number) is
\begin{equation}
\label{eq:transformation-u_kappa-u}
u_\kappa \equiv \frac{1}{\kappa} \textrm{arcsinh} (\kappa u)
= \ln [\exp_{\kappa} (u)],
\end{equation}
implies $d_\kappa u = d u_\kappa$, {\it i.e.},
the \textit{deformed differential} of an ordinary variable $u$
can be rewritten as with the ordinary differential of
a \textit{deformed variable} $u_\kappa$.
In this way, one defines the $\kappa$-derivative operator by
\begin{equation}
\label{eq:kappa-derivative}
  \displaystyle D_{\kappa} f(u)
        \equiv  \displaystyle
		  \lim_{u'\to u}\frac{f(u') - f(u)}{u'\overset{\kappa}{\ominus} u}
        = \displaystyle \sqrt{1+\kappa^2 u^2} \frac{df(u)}{du},
\end{equation}
with the $\kappa$-exponential
an eigenfunction of $D_\kappa$,
$D_\kappa \exp_\kappa u = \exp_\kappa u$.
Similarly, the dual $\kappa$-derivative is defined by
\begin{equation}
\label{eq:kappa-derivative-dual}
  \displaystyle \widetilde{D}_\kappa f(u)
        \equiv \displaystyle \lim_{u'\to u}\frac{f(u')\overset{\kappa}{\ominus} f(u)}{u'- u}
        = \displaystyle \frac{1}{\sqrt{1+\kappa^2 [f(u)]^2}} \frac{df(u)}{du},
\end{equation}
which satisfies $\widetilde{D}_\kappa \ln_\kappa u = 1/u$.
These operators obey
$\widetilde{D}_\kappa x(y) = [D_\kappa y(x)]^{-1}$.
In particular, we have
$D_{\kappa}u = (\widetilde{D}_\kappa u)^{-1} = \sqrt{1+\kappa^2 u^2}$.
From Eqs.~(\ref{eq:kappa-derivative}) and (\ref{eq:kappa-derivative-dual})
we see that the standard derivative is recovered as  $\kappa \rightarrow 0$.
The deformed derivative operator (\ref{eq:kappa-derivative}) can be seen as the variation
of the function $f(u)$ with respect to a nonlinear variation of the independent variable $u$,
{\it i.e}., $D_\kappa f(u) = df(u)/du_\kappa$.
On the other hand, the dual deformed derivative operator (\ref{eq:kappa-derivative-dual})
is the rate of change of a nonlinear variation of the function $f(u)$
with respect to the standard variation of the independent variable $u$,
$\widetilde{D}_\kappa f(u) = d_\kappa f(u)/du$.
The deformed  second derivatives satisfy
\begin{equation}
\label{eq:second-kappa-derivative}
  D_{\kappa}^2 f(u)
         = \sqrt{1+\kappa^2 u^2}
          \frac{d}{du}
             \left[
                   \sqrt{1+\kappa^2 u^2} \frac{df}{du}
             \right]
\end{equation}
and
\begin{equation}
\label{eq:second-kappa-derivative-dual}
  \widetilde{D}_{\kappa}^2 f(u)
		= \frac{1}{\sqrt{1+\kappa^2 [f(u)]^2}}
          \frac{d}{du}
             \left\{
                   \frac{1}{\sqrt{1+\kappa^2 [f(u)]^2}} \frac{df}{du}
             \right\}.
\end{equation}
These rules can be extended to deformed derivatives of higher order.

\section{\label{sec:kappa-deformed-dynamics}
		$\kappa$-Deformed dynamics of a system with position-dependent mass}

\subsection{$\kappa$-Deformed classical formalism}

Let us first consider the problem of a particle
with a position-dependent mass (PDM) $m(x)$ in 1D for
the classical formalism.
The Hamiltonian of the system is
\begin{equation}
\label{eq:hamiltonian_x_p}
 \mathcal{H}(x, p) = \frac{p^2}{2m(x)}+ V(x),
\end{equation}
whose the linear momentum is $p= m(x)\dot{x}$,
leads to the equation of motion
\begin{equation}
\label{eq:motion-equation-m(x)}
	m(x)\ddot{x}+\frac{1}{2}m'(x)\dot{x}^2 = F(x)
\end{equation}
with $F(x) = -dV/dx$ the force acting on the particle,
where $\dot{x} = dx/dt$, $\ddot{x} = d^2 x/dt^2$
and $m'(x) = dm/dx$ give velocity, acceleration and mass gradient,
respectively.
The point canonical transformation (PCT)
\begin{equation}
\label{eq:classical-dynamical-variables}
	\eta = \int^x \sqrt{\frac{m(y)}{m_0}}dy
\quad \textrm{and} \quad
	\Pi = \sqrt{\frac{m_0}{m(x)}} p,
\end{equation}
maps the Hamiltonian (\ref{eq:hamiltonian_x_p}) of a particle with
PDM $m(x)$ in the usual phase space $(x,p)$ into another
Hamiltonian of a particle with a constant mass
$m_0$ represented in the deformed phase space
$(\eta,\Pi)$,
\begin{equation}
	\mathcal{K}(\eta, \Pi) = \frac{1}{2m_0}\Pi^2 + U({\eta}),
\end{equation}
with $U(\eta) = V(x(\eta))$
the potential expressed in the deformed space-coordinate $\eta$.
When $m(x)=m_0$, both representations coincide.

Let us consider in particular the mass function
\begin{equation}
\label{eq:m(x)}
 m(x) = \frac{m_0}{1+\kappa^2 x^2}\,,
\end{equation}
where the parameter $\kappa$ has units of inverse length and
controls the dependence of the mass with position,
where $\kappa=0$ corresponds to the standard case.
Thus equation of motion (\ref{eq:motion-equation-m(x)}) becomes
\begin{equation}
 m_0 \left[ \frac{\ddot{x}}{(1+\kappa^2 x^2)}
			- \frac{\kappa^2 x \dot{x}^2}{(1+\kappa^2 x^2)^2}
		\right] = F(x).
\end{equation}
This equation can be compactly rewritten
in the form of a deformed Newton's second law
\begin{equation}
\label{eq:second_newton_law_generalized}
       m_0 \widetilde{D}^2_{\kappa} x (t) = F(x).
\end{equation}
Moreover, for the mass function (\ref{eq:m(x)})
the $\kappa$-deformed spatial coordinate and
its conjugated linear momentum are
\begin{subequations}
\label{eq:x_k-Pi_k}
\begin{equation}
\label{eq:kappa-pahse-space}
\eta = \frac{1}{\kappa} \textrm{arcsinh}(\kappa x)
\equiv x_\kappa,
\end{equation}
and
\begin{equation}
\Pi = \sqrt{1+\kappa^2 x^2} p
\equiv \Pi_\kappa,
\end{equation}
\end{subequations}
with Poisson brackets $\{ x_\kappa, \Pi_\kappa \}_{x,p} =1$.
The deformed displacement $d_\kappa x$ of a particle with the
non-constant mass $m(x)$, given in Eq.~(\ref{eq:m(x)}), is mapped into the usual
displacement $dx_\kappa$ in a deformed space $x_\kappa$
provided with a constant mass $m_0$:
$d_\kappa x \equiv
    (x+dx) \overset{\kappa}{\ominus} x
    = dx/\sqrt{1+\kappa^2 x^2}$,
up to first order.
The time evolution of the system is governed by the dual derivative, {\it i.e}.
$\widetilde{D}_{\kappa} x(t)
= \dot{x}/\sqrt{1+\kappa^2 x^2}$.

\subsection{$\kappa$-Deformed quantum formalism}

In the quantization of a PDM system an ordering ambiguity arises
for defining the kinetic energy operator in terms of
the mass operator $m(\hat{x})$ and the linear momentum $\hat{p}$.
There are several ways to define a Hermitian kinetic energy operator,
and a general two-parameter form is given by
\begin{equation}
\label{eq:general-kinetic-operator-pdm}
\hat{T} =
    \frac{1}{4} \left\{ [m(\hat{x})]^{-\alpha} \hat{p}
	[m(\hat{x})]^{-1+\alpha+\beta} \hat{p} [m(\hat{x})]^{-\beta}
	+[m(\hat{x})]^{-\beta} \, \hat{p}
	[m(\hat{x})]^{-1+\alpha+\beta} \hat{p} [m(\hat{x})]^{-\alpha}
	\right\}.
\end{equation}
For more details see the discussions, for instance, of
von Roos,
\cite{vonroos_1983}
L\'evy-Leblond,
\cite{Levy-Leblond-1995}
and others.
Among many particular cases in the literature, we point out
the proposals by
Ben Daniel and Duke
($\alpha = \beta = 0$),
\cite{BenDaniel-Duke-1966}
Gora and Williams
($\alpha = 1, \beta = 0$),
\cite{Gora-Williams-1969}
Zhu and Kroemer
($\alpha = \beta = \frac{1}{2}$),
\cite{Zhu-Kroemer-1983}
Li and Kuhn
($\alpha = \frac{1}{2}, \beta = 0$).
\cite{Li-Kuhn-1993}
Morrow and Brownstein
\cite{Morrow-Brownstein-1984}
have shown that only the case $\alpha = \beta$ satisfies
the conditions of continuity of the wave-function
at the boundaries of a heterojunction in crystals.
In particular, Mustafa and Mazharimousavi
\cite{Mustafa-Mazharimousavi-2007}
have shown that the case $\alpha = \beta = \frac{1}{4}$
allows the mapping of a quantum Hamiltonian with PDM
into a Hamiltonian with constant mass
by means a PCT.
More precisely, considering the quantum Hamiltonian
\begin{equation}
\label{eq:hamiltonian-MM}
 \hat{H}(\hat{x}, \hat{p}) = \frac{1}{2}
	[m(\hat{x})]^{-\frac{1}{4}}
	\hat{p} [m(\hat{x})]^{-\frac{1}{2}} \hat{p}
	[m(\hat{x})]^{-\frac{1}{4}}
 	+ V(\hat{x}),
\end{equation}
the Schr\"odinger equation
$i\hbar \frac{\partial}{\partial t}| \Psi \rangle
 = \hat{H}| \Psi \rangle$
in the position representation $\{ |\hat{x}\rangle \}$ reads
\begin{equation}
\label{eq:SE-MM-m(x)}
 i\hbar \frac{\partial \Psi (x, t)}{\partial t}
 = \left( -\frac{\hbar^2}{2m_0} \sqrt[4]{\frac{m_0}{m(x)}}
 \frac{\partial}{\partial x} \sqrt{\frac{m_0}{m(x)}}
 \frac{\partial}{\partial x}\sqrt[4]{\frac{m_0}{m(x)}}
 + V(x) \right) \Psi (x, t),
\end{equation}
with $\Psi (x, t)=\psi(x)e^{-iEt/\hbar}$
and $E$ the eigenvalue corresponding to
the eigenfunction $\psi(x)$ of $\hat{H}$.
It is straightforwardly verified that
the probability density $\rho(x,t) \equiv  |\Psi(x,t)|^2$
satisfies the continuity equation
\begin{equation}
\label{eq:contituity-equation-pdm-sytem}
	\frac{\partial \rho (x,t)}{\partial t} =
	-\frac{\partial J(x,t)}{\partial x},
\end{equation}
where the probability current is
\begin{equation}
\label{eq:current-density}
	J(x,t) \equiv
		\operatorname{Re}
		\left\{
		\Psi^{\ast} (x,t) \left( \frac{\hbar}{i}
		\frac{\partial}{\partial x} \right)
		\left[\frac{1}{m(x)} \Psi (x,t)\right]\,
		\right\}.
\end{equation}
Equation~(\ref{eq:SE-MM-m(x)}) can be conveniently rewritten
by means of the transformation
$\Psi (x, t) = \sqrt[4]{m(x)/m_0} \Phi (x, t)$,
as
\begin{equation}
\label{eq:m(x)-SE}
i\hbar \frac{\partial \Phi (x, t)}{\partial t} =
\left[ -\frac{\hbar^2}{2m_0} \left( \sqrt{{\frac{m_0}{m(x)}}}
\frac{\partial}{\partial x} \right)^2 +V(x)\right] \Phi (x, t).
\end{equation}

Let us consider in particular the mass function (\ref{eq:m(x)}).
The modified wave-function
$\Phi (x, t) =  \sqrt[4]{1+\kappa^2 x^2} \Psi (x, t)$
obeys a {\it $\kappa$-deformed Schr\"odinger wave-equation}
\begin{equation}
\label{eq:kappa-SE}
i\hbar \frac{\partial \Phi (x, t)}{\partial t} =
-\frac{\hbar^2}{2m_0} \mathcal{D}_{\kappa}^2 \Phi (x, t)
+V(x) \Phi (x, t)
\end{equation}
with
$\mathcal{D}_{\kappa} = \sqrt{1+\kappa^2 x^2}\partial_x$,
which is the analog of the $\kappa$-derivative operator
(\ref{eq:kappa-derivative}).
Using Eq.~(\ref{eq:second-kappa-derivative-dual}), we obtain
\begin{equation}
\label{eq:kappa-SE-explicitly}
i\hbar \frac{\partial \Phi (x, t)}{\partial t} =
-\frac{\hbar^2 (1+\kappa^2 x^2)}{2 m_0}
\frac{\partial^2 \Phi (x, t)}{\partial x^2}
-\frac{\hbar^2 \kappa^2 x}{2 m_0} \frac{\partial \Phi (x, t)}{\partial x}
+ V(x) \Phi (x, t).
\end{equation}
Equation~(\ref{eq:kappa-SE})
is indeed equivalent to a Schr\"odinger-like equation for
$\Phi (x,t)$ with the non-Hermitian Hamiltonian operator
\begin{equation}
\label{eq:H_kappa}
\hat{H}_\kappa = \frac{1}{2m_0}\hat{p}_\kappa^2 + V(\hat{x}),
\end{equation}
where
$\hat{p}_\kappa \equiv -i\hbar \mathcal{D}_{\kappa}
= \sqrt{1+\kappa^2 \hat{x}^2}\hat{p}$
stands for a $\kappa$-deformed non-Hermitian momentum operator,
and obeys the commutation relation
\begin{equation}
[\hat{x}, \hat{p}_\kappa] =
i\hbar\sqrt{\hat{1} + \kappa^2 \hat{x}^2}.
\end{equation}
This leads to generalized uncertainty principle
$\Delta x \Delta p_\kappa \geq \frac{\hbar}{2}
\langle \sqrt{ 1+\kappa \hat{x}^2 } \rangle$.
We notice that if the standard wave-function
$\Psi(x, t)$ is normalized, then $\Phi (x, t)$ is normalized under a
$\kappa$-deformed integral. Indeed, we have
\begin{equation}
\label{eq:normalization}
\int_{x_{i}}^{x_f} \Psi^{\ast} (x, t) \Psi (x, t) dx =
\int_{x_i}^{x_f} {\Phi}^{\ast} (x, t) \Phi (x, t) d_\kappa x = 1.
\end{equation}
Besides, we obtain
the $\kappa$-deformed continuity equation
\begin{equation}
\label{eq:deformed-contituity-equation}
  \frac{\partial \varrho(x,t)}{\partial t}
  + \mathcal{D}_{\kappa} \mathcal{J} (x, t) = 0,
\end{equation}
with $\varrho (x, t) = |\Phi(x, t)|^2$ and
\begin{equation}
 \label{eq:deformed-current-density}
	\mathcal{J} (x, t) \equiv
		\operatorname{Re}
		\left\{
		\Phi^{\ast} (x, t)
		\left( \frac{\hbar}{i}
		\mathcal{D}_{\kappa}\right)
		\left[ \frac{\Phi (x, t)}{m_0} \right]\,
		\right\}.
\end{equation}

It is worth noting that there is an equivalence between
the Schr\"odinger equation for the Hermitian system
(\ref{eq:hamiltonian-MM})
with the mass function $m(x)$ given by (\ref{eq:m(x)}) and the
non-Hermitian one (\ref{eq:H_kappa}) expressed in terms of
a $\kappa$-deformed momentum operator, where $\Psi(x, t)$
must be replaced by
$\Phi (x, t) =  \sqrt[4]{1+\kappa^2 x^2} \Psi (x, t)$.
Moreover, we see that in the description of quantum
systems with the mass function (\ref{eq:m(x)}) in terms of the
modified wave-function $\Phi (x,t)$,
the usual derivative and integral with respect to
the variable $x$ are replaced by
their corresponding $\kappa$-deformed versions.
Analogous features apply in the classical formalism, with
the motion equation expressed in terms of the dual
$\kappa$-derivative
(see Eq.~(\ref{eq:second_newton_law_generalized})).

Using the change of variable
$x\rightarrow x_\kappa =\ln[\exp_\kappa(x)]$
(see Eq.~(\ref{eq:transformation-u_kappa-u})),
then Eq.~(\ref{eq:kappa-SE}) can be rewritten
in the $\kappa$-deformed space as
\begin{equation}
\label{eq:SE-deformed-space}
i\hbar \frac{\partial \Lambda  (x_{\kappa}, t)}{\partial t} =
-\frac{\hbar^2}{2m_0} \frac{\partial^2 \Lambda (x_{\kappa}, t)}{\partial x_{\kappa}^2}
+U(x_{\kappa}) \Lambda  (x_{\kappa}, t),
\end{equation}
with $\Lambda (x_\kappa, t) = \Phi (x(x_\kappa), t)$
and $U(x_\kappa)=V(x(x_\kappa))$ a modified potential in terms
of the original one $V$ and the inverse transformation $x=x(x_\kappa)$.
Therefore, the wave-equation for
$\Psi (x,t)$ of a system with PDM (\ref{eq:m(x)})
with the potential $V(x)$ in the standard space $\{ | x \rangle \}$
is mapped into an equation for
$\Lambda  (x_\kappa,t)$ with the potential $U(x_\kappa)=V(x(x_\kappa))$
in the deformed space $\{|\hat{x}_\kappa \rangle\} $.
The quantum Hamiltonian associated with the Schr\"odinger
wave-equation (\ref{eq:SE-deformed-space}) is
$\hat{K}(\hat{x}_\kappa,\hat{\Pi}_\kappa)
 = \frac{1}{2m_0}\hat{\Pi}_\kappa^2 + U(\hat{x}_\kappa)$,
that can be obtained by
applying the point canonical
transformation
$(\hat{x}, \hat{p}) \rightarrow
(\hat{x}_\kappa, \hat{\Pi}_\kappa)$
on the quantum Hamiltonian (\ref{eq:hamiltonian-MM})
where
\begin{subequations}
\label{eq:hat-x_k-Pi_k}
\begin{align}
\hat{x}_\kappa &=
\frac{1}{\kappa} \textrm{arcsinh}(\kappa \hat{x}),
\\
\hat{\Pi}_\kappa &= \sqrt[4]{1+\kappa^2 \hat{x}^2}\,\hat{p}\,
                 \sqrt[4]{1+\kappa^2 \hat{x}^2}
			   = \frac{1}{2} (\hat{p}_\kappa^{\dagger}+\hat{p}_\kappa),
\end{align}
\end{subequations}
with $[\hat{x}_\kappa, \hat{\Pi}_\kappa] =i \hbar\hat{1}$.
Also, we have that $\hat{\Pi}_\kappa$
is in accordance with the definition of a
PDM pseudo-momentum operator introduced in
Ref.~\onlinecite{Mustafa-Mazharimousavi-2007}.
Thus, the dynamical variables
(\ref{eq:classical-dynamical-variables})
are the classical counterparts of the Hermitian operators
(\ref{eq:hat-x_k-Pi_k}).

From the eigenvalue equation
$\hat{\Pi}_\kappa |k\rangle = \hbar k |k\rangle $,
the eigenfunctions in the representation $\{ |\hat{x} \rangle \}$
result
\begin{align}
\psi_k (x)
&= \frac{C}{\sqrt[4]{1+\kappa^2 x^2}} [\exp_\kappa (x)]^{ik}
\nonumber \\
&= \frac{C}{\sqrt[4]{1+\kappa^2 x^2}}
    \exp \left[ \frac{ik}{\kappa} \textrm{arcsinh} (\kappa x) \right],
\end{align}
where $C$ is a constant.
As in the non deformed case ($\kappa=0$), the function
$\psi_k(x)$ is not normalizable.
Even though, a deformed wave-packet can be defined
from the $\kappa$-deformed Fourier transform
\cite{Scarfone-2017}
\begin{equation}
\label{eq:Fourier-transform}
\psi (x) = \frac{1}{\sqrt[4]{1+\kappa^2 x^2}}
			  \int_{-\infty}^{+\infty} g(k)
			  e^{ \frac{ik}{\kappa} \textrm{arcsinh} (\kappa x) } dk,
\end{equation}
where
$g(k)$
is the distribution function of the wave-vectors $k$.
It is verified straightforwardly that the corresponding wave-packet
of the operator $\hat{p}_\kappa$ is
$\varphi(x) = \int_{-\infty}^{+\infty} g(k) [\exp_\kappa (x)]^{ik} dk$.
The wave-packet in the representation of the deformed space
is
$\phi (x_\kappa) = \varphi(x(x_\kappa)) =
\sqrt[4]{1+\kappa^2 x^2} \psi(x(x_\kappa))
=  \int_{-\infty}^{+\infty} g(k) e^{ik x_\kappa}dk$.
From the Plancherel theorem, we have
\begin{align}
\label{eq:inverse-Fourier-transform}
g(k) &= \frac{1}{2\pi} \int_{-\infty}^{+\infty} \phi(x_\kappa) e^{-ik x_\kappa}dx_\kappa
	 \nonumber \\
     &= \frac{1}{2\pi} \int_{-\infty}^{+\infty} \varphi(x) [\exp_\kappa (x)]^{-ik} d_\kappa x
     \nonumber \\
     &= \frac{1}{2\pi} \int_{-\infty}^{+\infty} \frac{\psi(x)}{\sqrt[4]{1+\kappa^2 x^2}}
		 e^{-\frac{ik}{\kappa} \textrm{arcsinh} (\kappa x) } dx.
\end{align}

\section{\label{sec:particle-infinite-well}
		 Particle in an infinite potential well}

In Secs.~\ref{sec:particle-infinite-well} and
\ref{sec:kappa-deformed-oscillator} we illustrate the quantum and classical
$\kappa$-deformed formalism with two paradigmatic examples.

\subsection{Classical case}

First we consider
the problem of a particle confined in an infinite square potential well
between $x = 0$ and $x = L$.
If
$\mathcal{H}(x, p) = E$
is the energy of the
classical particle, then the linear momentum is
$p(x) =\pm \sqrt{2m_0E/(1+\kappa^2 x^2)}$
and the velocity is
$v(x) = \pm v_0 \sqrt{1+\kappa^2 x^2}$ with
$v_0 = \sqrt{2E/m_0}$.
For $v(0)=v_0$ and $0<x<L$, the position as a
function of time is
$x(t) = \ln_\kappa [\exp(v_0 t)]$.
Hence, the classical probability density
$\rho_{\textrm{classic}}(x)dx \propto dx/v$
to find the particle within the interval $[x, x+dx]$ is
\begin{equation}
\label{eq:rho_classic}
\rho_{\textrm{classic}}(x)dx
= \frac{\kappa }{\ln \left( \kappa L + \sqrt{1+\kappa^2 L^2} \right)}
\frac{dx}{\sqrt{1+\kappa^2 x^2}},
\end{equation}
from which the uniform
distribution
$\rho_{\textrm{classic}}(x) = 1/L$
is recovered when $\kappa \rightarrow 0$.
The first and the second moments of the position and the linear
momentum for the classical distribution (\ref{eq:rho_classic}) are
\begin{subequations}
\label{eq:classic_first_second_moment}
\begin{align}
\label{eq:x-med-classical}
&\frac{\overline{x}}{L} =
\frac{\sqrt{1 + \kappa^2 L^2} - 1}{
\kappa L \ln \left( \kappa L + \sqrt{1+\kappa^2 L^2} \right)},
\\
\label{eq:x^2-med-classical}
&\frac{\overline{x^2}}{L^2} =
\frac{1}{2\kappa^2 L^2} \left[
\frac{\kappa L \sqrt{1+\kappa^2 L^2}}{
\ln \! \left( \kappa L + \sqrt{1+\kappa^2 L^2} \right)} -1
\right],
\\
\label{eq:p-med}
&\overline{p} = 0,
\\
\label{eq:p^2-med}
&\overline{p^2} = 2m_0 E \left[
\frac{\kappa L}{\sqrt{1+\kappa^2 L^2}
\ln \! \left( \kappa L + \sqrt{1+\kappa^2 L^2} \right)}
\right].
\end{align}
\end{subequations}
We can verify that
$\lim_{\kappa \rightarrow 0}\overline{x} = L/2$,
$\lim_{\kappa \rightarrow 0}\overline{x^2} = L^2/3$
and
$\lim_{\kappa \rightarrow 0}\overline{p^2} = 2m_{0}E$.
From the change of variable $x \rightarrow x_\kappa$
the PDM particle confined in an interval $[0,L]$ is mapped into a
particle with constant mass in $[0,L_\kappa]$, where
$L_\kappa = \textrm{arcsinh}(\kappa L)/\kappa$
corresponds to the length of the box in the deformed space.

\subsection{Quantum case}

\begin{figure}[!h]
\centering
\includegraphics[width=0.38\linewidth]{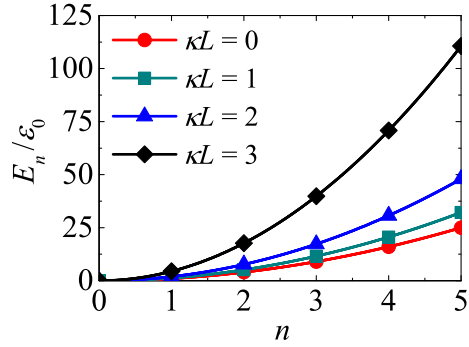}
\caption{\label{fig:1}
(Color online)
Energy levels of a particle with PDM
$m(x) = m_0/(1+\kappa^2 x^2)$ in an
infinite square well of size $L$,
for different quantum numbers $n$ and values of $\kappa$,
given in terms of the nondeformed fundamental energy
$\varepsilon_0 = \frac{\hbar^2 \pi^2}{2m_0L^2}$.
The values of the energies are discrete, and the solid lines
help for guiding the eyes.
}
\end{figure}

Let us now analyze the problem
in the $\kappa$-deformed quantum formalism.
Considering
$\Phi (x, t) = \varphi (x) e^{-iEt/\hbar},$
this leads to the time-independent
Schr\"odinger-like equation
$
-\frac{\hbar^2}{2m_0} D_{\kappa}^2
\varphi (x) = E \varphi (x),
$
whose eigenfunctions are given by
\begin{equation}
\label{eq:psi_n}
\varphi_n (x) = \sqrt[4]{1+ \kappa^2 x^2} \psi_n(x) = C_\kappa
\sin \left[ \frac{k_{\kappa,n}}{\kappa}
\textrm{arcsinh} (\kappa x) \right]
\end{equation}
for $0 \leq x \leq L$, and $\varphi_n(x) = 0$ elsewhere,
with $C_{\kappa}^2 = 2/L_\kappa$ and
$k_{\kappa,n} =  n \pi/L_\kappa$, where $n$ is an integer number
and $L_\kappa = \kappa^{-1} \textrm{arcsinh} (\kappa L)$.
The energy levels corresponding to these eigenfunctions are
\begin{equation}
\label{eq:energy}
E_n =
\frac{\hbar^2 \pi^2 n^2 \kappa^2}{2m_0\textrm{arcsinh}^2 (\kappa L)}
= \varepsilon_0 \left[
\frac{ \kappa L}{\textrm{arcsinh} (\kappa L)}
\right]^{2} n^2
\end{equation}
with $\varepsilon_0 = \hbar^2 \pi^2/(2m_0L^2)$.
The effect of the deformation parameter $\kappa$
corresponds to a contraction of the space
($L_\kappa <L $ for $\kappa \neq 0 $),
and consequently this leads to
an increase of the energy levels of the particle.
In Fig.~\ref{fig:1} we illustrate the energy levels of the
particle as a function of the quantum number for different
values of $\kappa$.

\begin{figure*}[!ht]
\centering
\begin{minipage}[h]{0.31\linewidth}
\includegraphics[width=\linewidth]{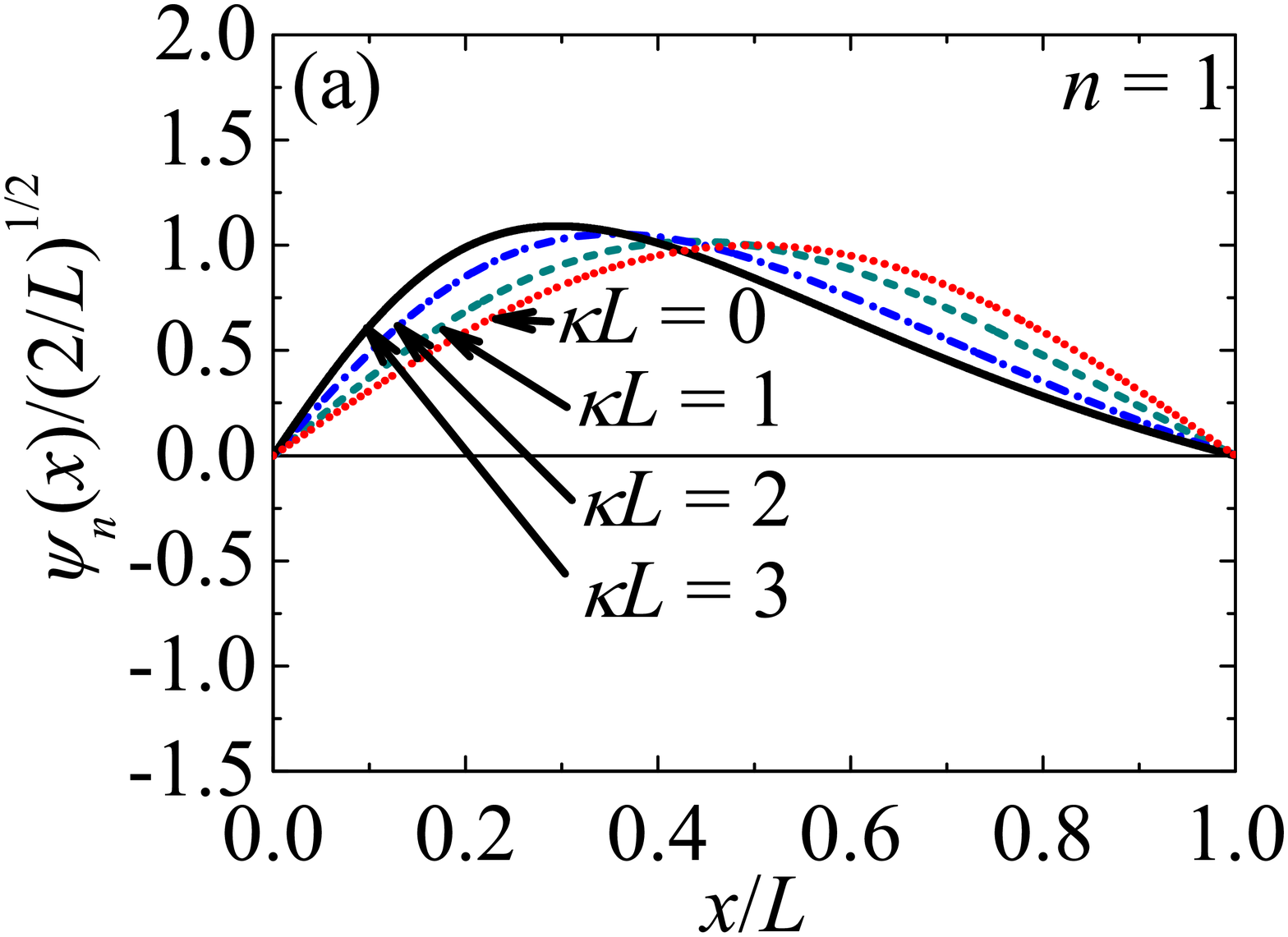}
\end{minipage}
\begin{minipage}[h]{0.31\linewidth}
\includegraphics[width=\linewidth]{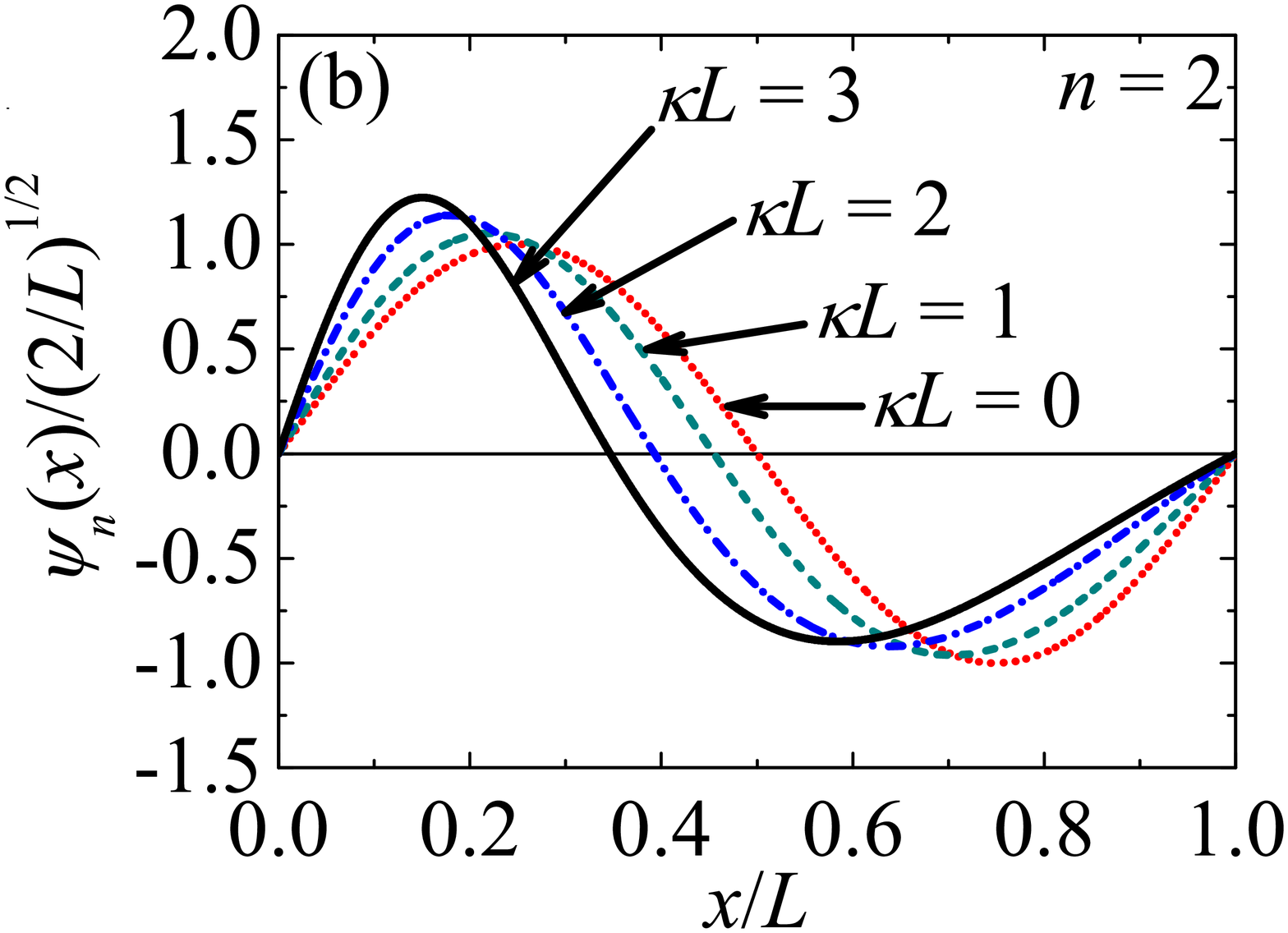}
\end{minipage}
\begin{minipage}[h]{0.31\linewidth}
\includegraphics[width=\linewidth]{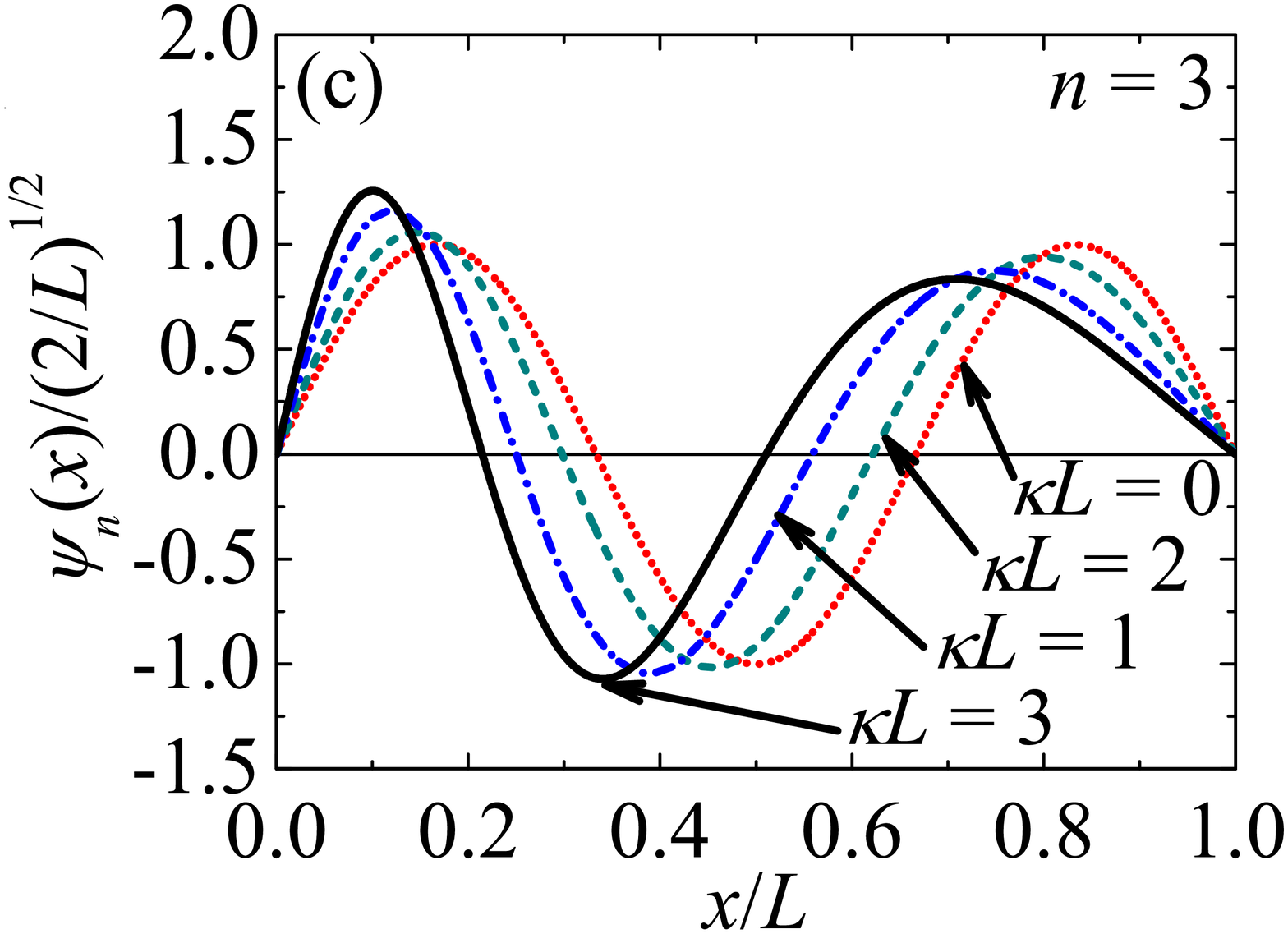}
\end{minipage} \\
\begin{minipage}[h]{0.31\linewidth}
\includegraphics[width=\linewidth]{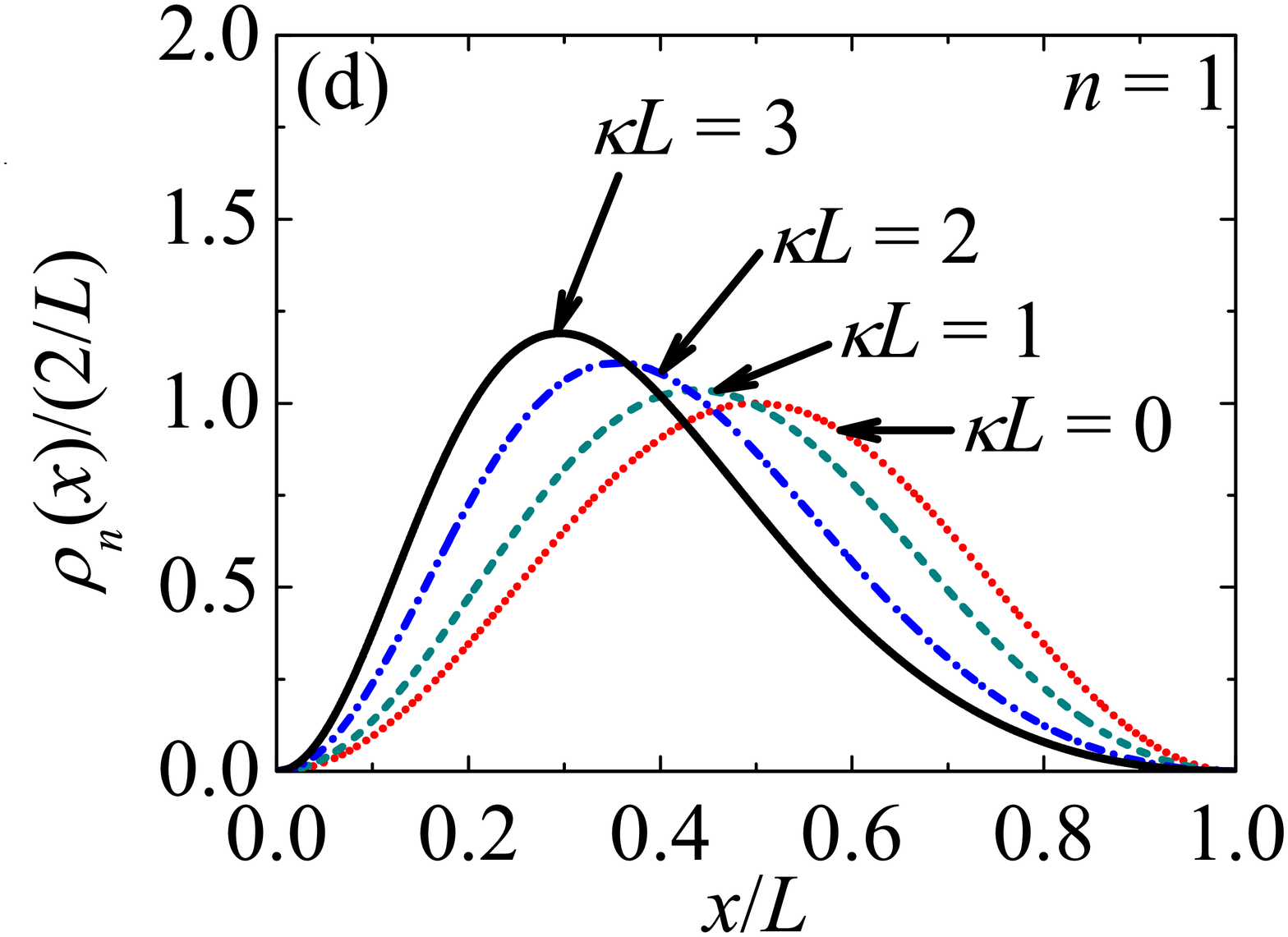}
\end{minipage}
\begin{minipage}[h]{0.31\linewidth}
\includegraphics[width=\linewidth]{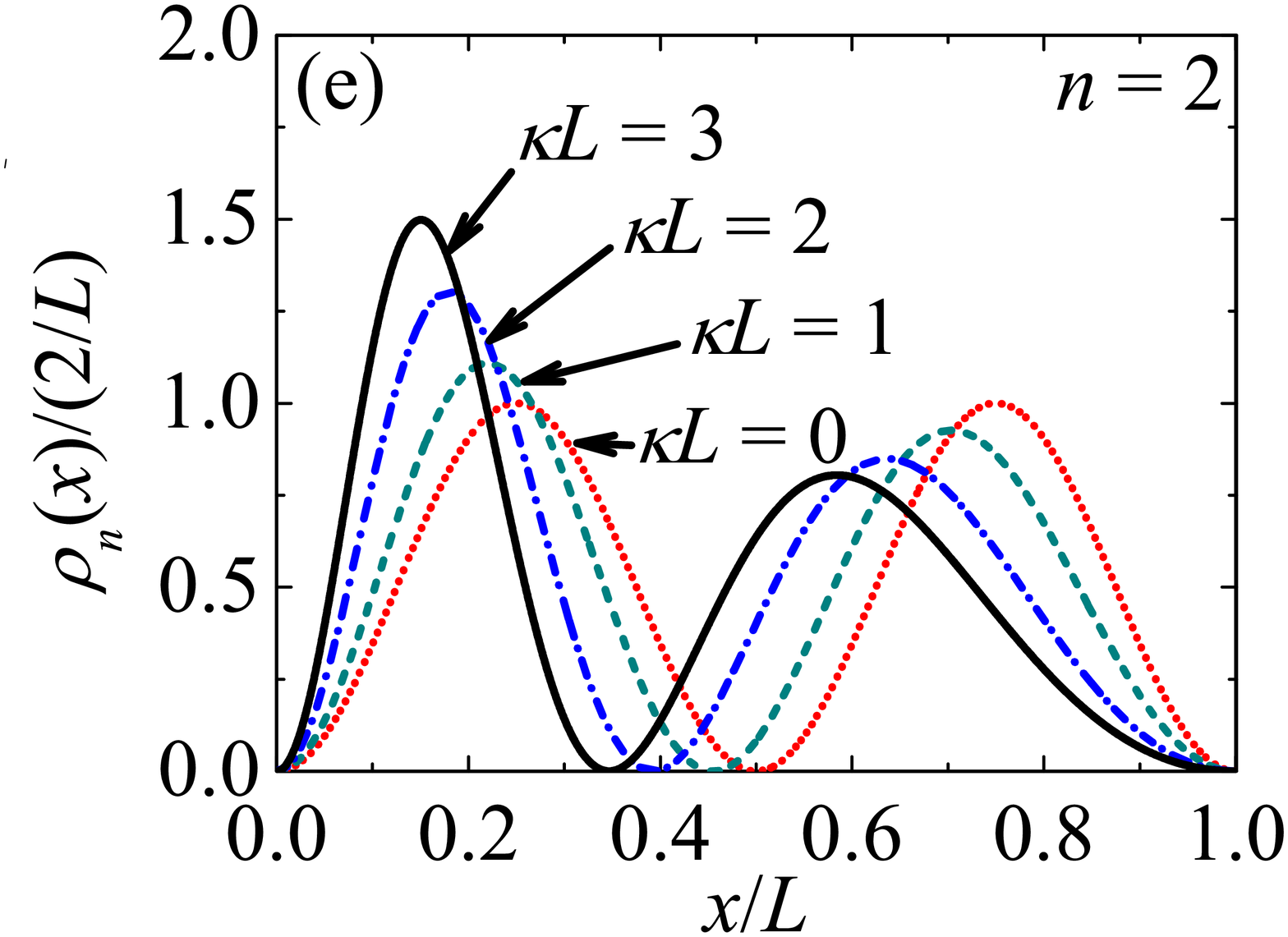}
\end{minipage}
\begin{minipage}[h]{0.31\linewidth}
\includegraphics[width=\linewidth]{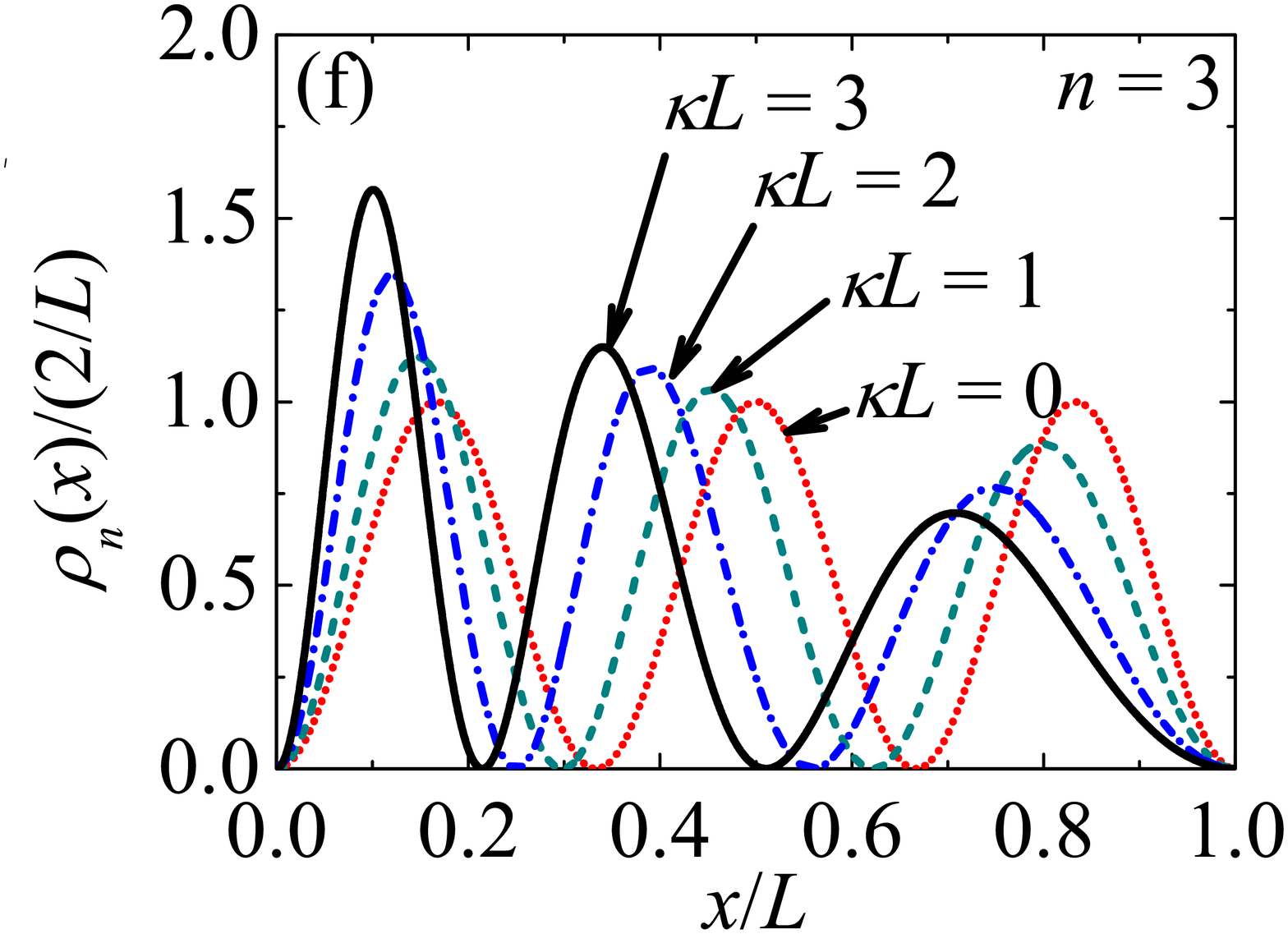}
\end{minipage} \\
\begin{minipage}[h]{0.31\linewidth}
\includegraphics[width=\linewidth]{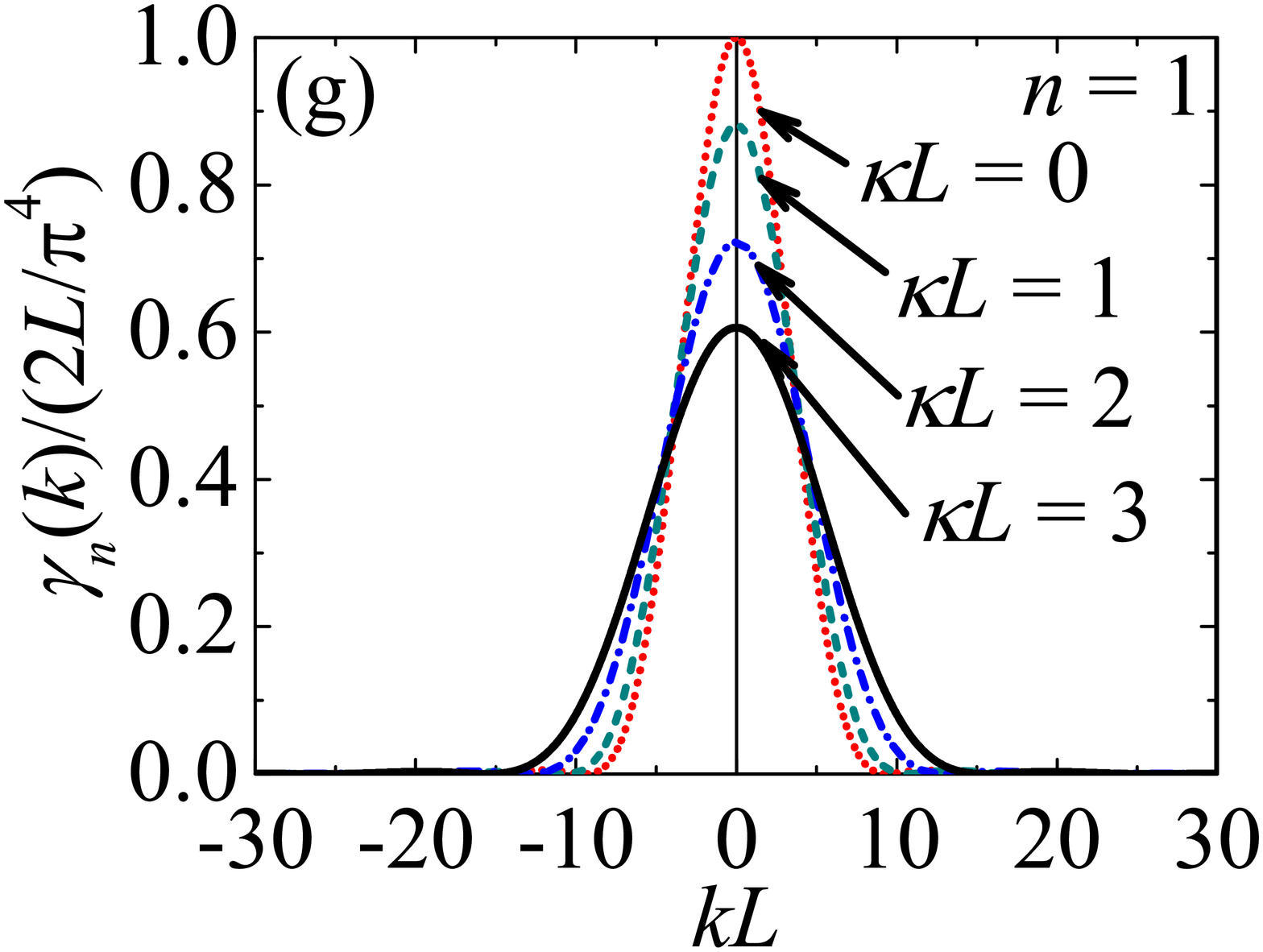}
\end{minipage}
\begin{minipage}[h]{0.31\linewidth}
\includegraphics[width=\linewidth]{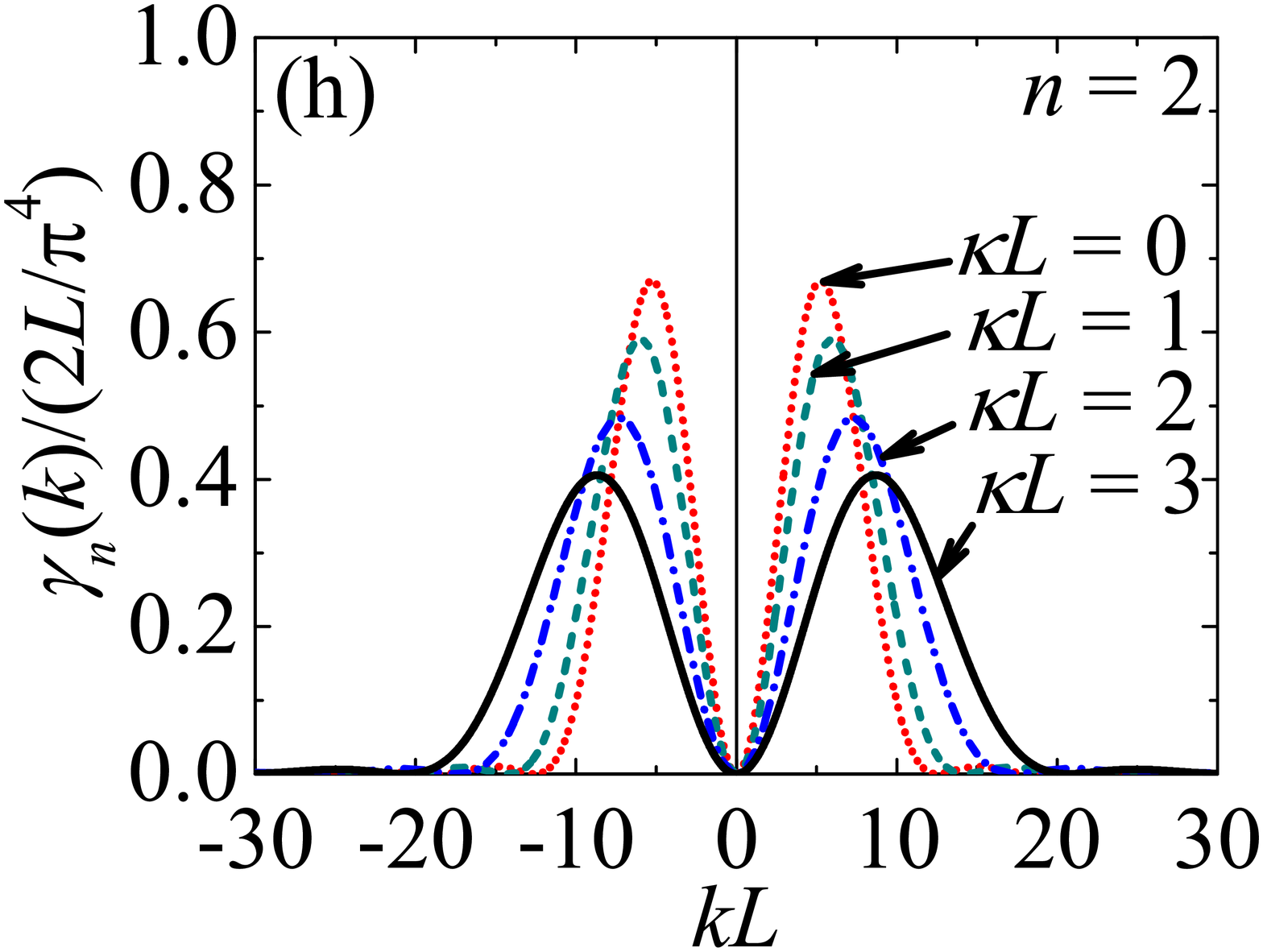}
\end{minipage}
\begin{minipage}[h]{0.31\linewidth}
\includegraphics[width=\linewidth]{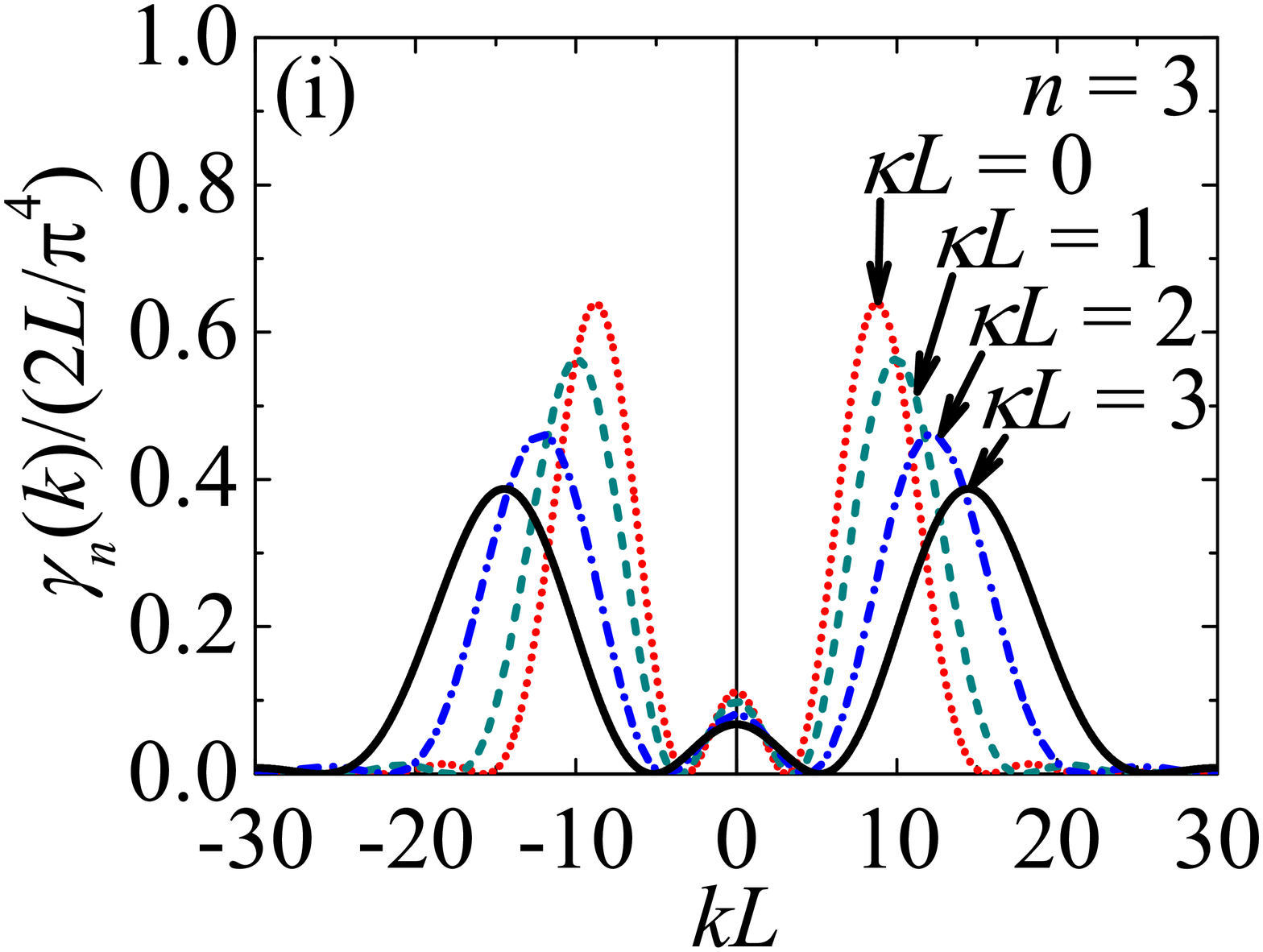}
\end{minipage}
\caption{\label{fig:2}
(Color online)
Eigenfunctions $\psi_n (x)$ ((a)-(c)),
probability densities  $\rho_n (x)=|\psi_n(x)|^2$
((d)-(f)) and $\gamma_n (k)=|g_n(k)|^2$ ((g)-(i))
for a particle with PDM $m(x)=m_0/(1+\kappa^2 x^2)$
and confined in an infinite square well
for different parameters $\kappa L$ (the usual case, $\kappa L=0$,
is shown for comparison).
[(a), (d) and (g)] $n=1$ (ground state),
[(b), (e) and (h)] $n=2$ (first excited state),
[(c), (f) and (i)] $n=3$ (second excited state).
}
\end{figure*}

\begin{figure}[!htb]
\centering
\begin{minipage}[h]{0.38\linewidth}
\includegraphics[width=\linewidth]{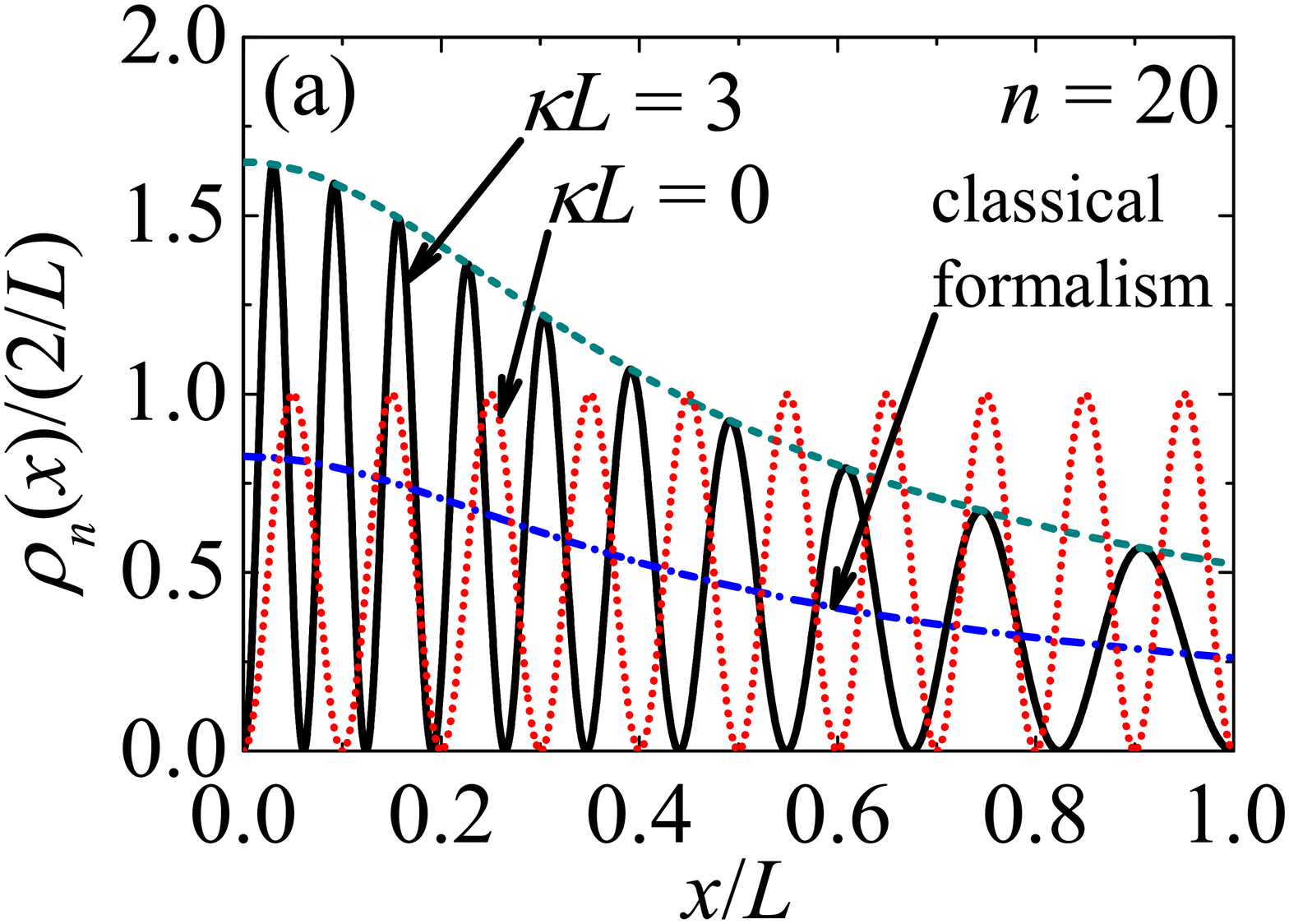}
\end{minipage}
\begin{minipage}[h]{0.38\linewidth}
\includegraphics[width=\linewidth]{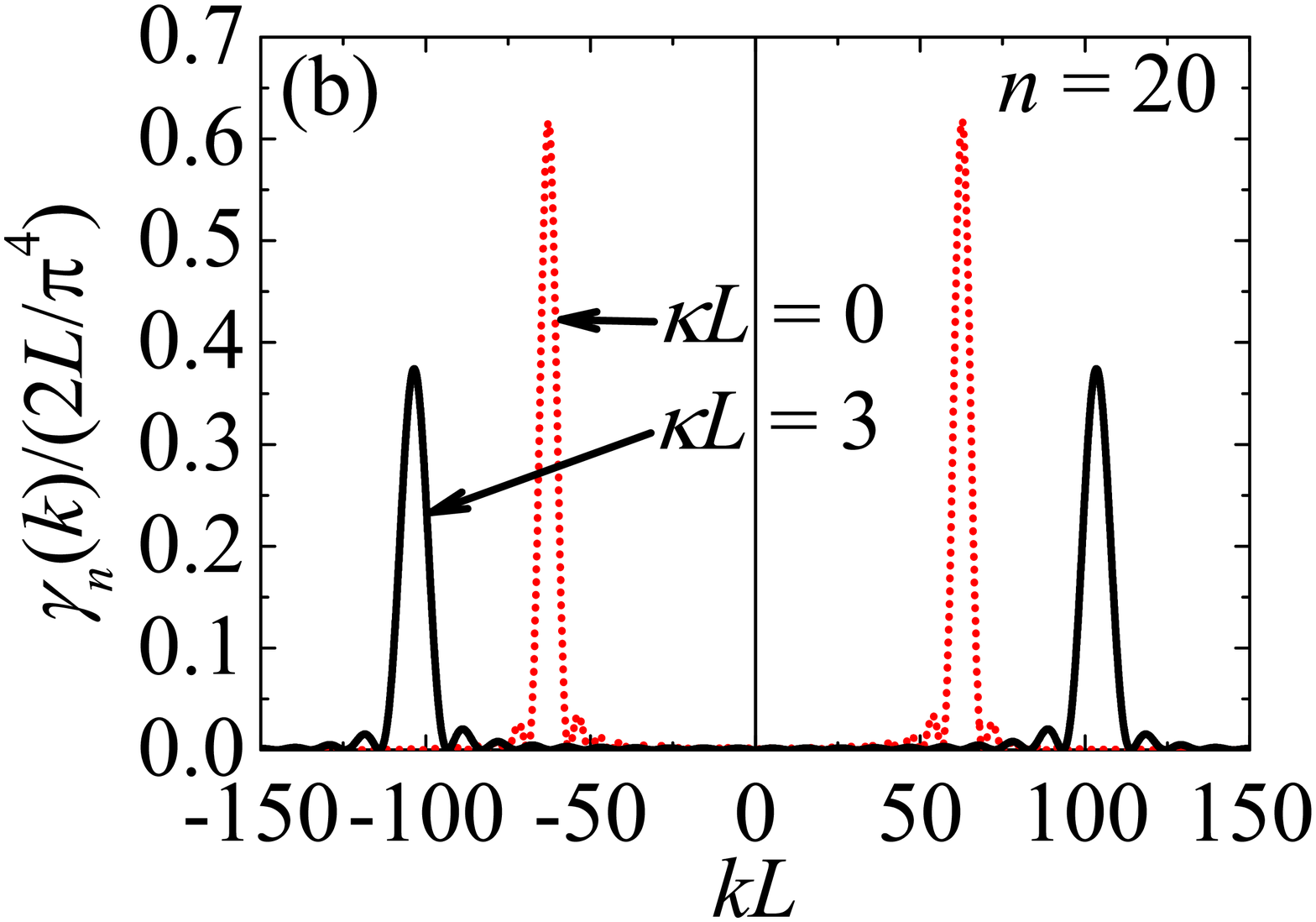}
\end{minipage}
\caption{\label{fig:3}
(Color online)
Probability densities (a) $\rho_n (x)=|\psi_n(x)|^2$
and (b) $\gamma_n (k) =|g_n(k)|^2$ of a particle with PDM
confined in an infinite square well for $\kappa L = 3.0$
and for the eigenstate $n=20$.
In the panel (a), the classical distribution
[Eq.~(\ref{eq:rho_classic})] is shown for
comparison, and the dotted upper line is
$2{\kappa}L/[\textrm{arcsinh} (\kappa x)
\sqrt{1+\kappa^2 x^2}]$.
}
\end{figure}

The probability densities of the stationary states in position space are
\begin{equation}
\label{eq:rho_n}
\rho_n (x)
	= |\psi_n (x)|^2
	= \frac{2\kappa}{\textrm{arcsinh} (\kappa L)}
		\frac{1}{\sqrt{1+ \kappa^2 x^2}}
		\sin^2 \left[ \frac{k_{\kappa,n}}{\kappa}
		\textrm{arcsinh} (\kappa x) \right].
\end{equation}
Substituting Eq.~(\ref{eq:psi_n}) into the
inverse Fourier transform (\ref{eq:inverse-Fourier-transform}),
we obtain the eigenfunctions for the particle confined in a box
in momentum space $k$
\begin{equation}
\label{eq:g_n(k)}
g_n(k) = n\sqrt{\frac{L_{\kappa}}{2}}
\left[ \frac{1+(-1)^{n+1}
e^{-ikL}}{(kL_{\kappa})^2 - (n{\pi})^2} \right].
\end{equation}
Consequently, its associated probability density results
\begin{align}
\label{eq:gamma_n(k)}
\gamma_n(k) = |g_n(k)|^2
= n^2L_\kappa
\frac{1-\cos (n{\pi}) \cos (kL_{\kappa})}{[(kL_{\kappa})^2-(n{\pi})^2]^2}.
\end{align}
Interestingly, the eigenfunctions (\ref{eq:g_n(k)}) and
the probability densities (\ref{eq:gamma_n(k)})
have the same form as in the case of
a particle with constant mass,
but with $L_\kappa$ instead of $L$.
In Fig.~\ref{fig:2} we plot the eigenfunctions $\psi_n(x)$ and
their probability densities in the coordinate and momentum spaces,
$\rho_n(x)$ and $\gamma_n(k)$,
for the three states of lower energy and for some values of
the deformation parameter $\kappa$.
We can see that as $\kappa$ increases,
$\rho_n(x)$ becomes more asymmetric and $\gamma_n(k)$
more spread along its domain.
In Fig.~\ref{fig:3} we show that the average value of the
quantum probability density $\rho_n(x)$
approaches to the classical probability density
$\rho_{\textrm{classic}}(x)$ (illustrated here for $n = 20$)
in accordance with the correspondence principle.
The distribution $\gamma_n(k)$
is also shown for the same state $n=20$.

\begin{figure}[!htb]
\centering
\begin{minipage}[h]{0.38\linewidth}
\includegraphics[width=\linewidth]{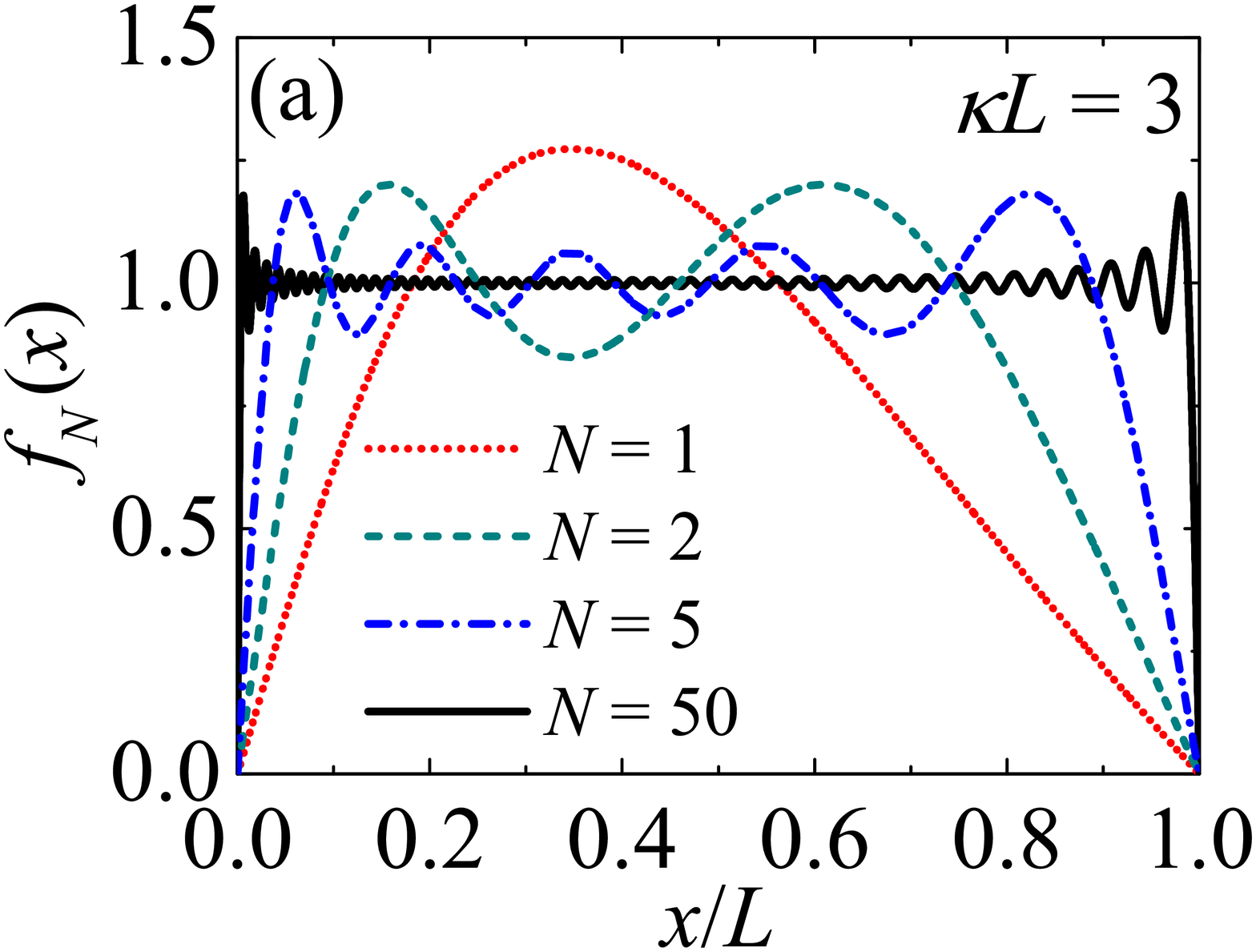}
\end{minipage}
\begin{minipage}[h]{0.38\linewidth}
\includegraphics[width=\linewidth]{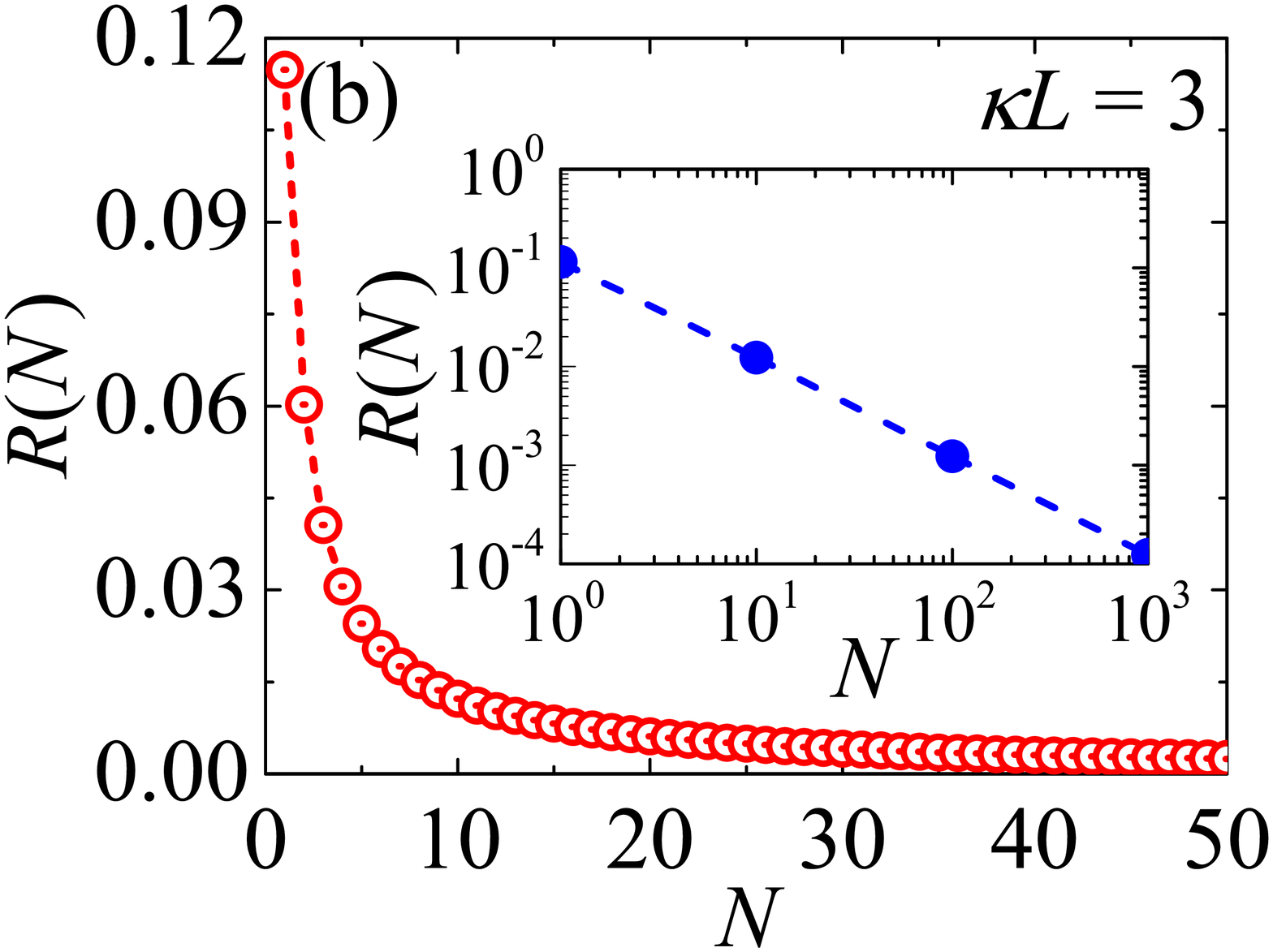}
\end{minipage}
\caption{\label{fig:4}
(Color online)
(a) Partial sum $f_N(x)$ ($\kappa$-deformed Fourier series)
[Eq.(\ref{eq:f_N(x)})] for $N=1,2,5$ and $50$.
(b) Mean square error of the approximation $R(N)$
for the range from $N$ = 1 to 50
(in the log-log graph inset: $N = 1$ to $10^3$).
}
\end{figure}

The eigenfunctions (\ref{eq:psi_n}) constitute
an orthonormal set of functions
that obey the inner product
$\int_0^{L} \varphi_n (x) \varphi_{n'}(x) d_\kappa x = \delta_{n,n'}$,
so that any continuous function in the interval
$[0,L]$ can be written as a linear combination
\begin{equation}
\label{eq:kappa-Fourier-series}
f(x) = \sum_{n=1}^{\infty}
c_n \sin \left[ n \pi
\frac{\textrm{arcsinh} (\kappa x)}{\textrm{arcsinh} (\kappa L)}
\right],
\end{equation}
with the coefficients $c_n$
of the series given by
\begin{equation}
c_n = \frac{2\kappa}{\textrm{arcsinh} (\kappa L)}
\int_0^{L} f(x) \sin \left[ n \pi
\frac{\textrm{arcsinh} (\kappa x)}{\textrm{arcsinh} (\kappa L)}
\right] d_\kappa x.
\end{equation}
Concerning the Sturm-Liouville problem,
Braga {\it et al.}
\cite{Braga-CostaFilho-2016}
have introduced a Fourier series
in terms of deformed trigonometric functions that
emerge from the formalism studied in
Ref.~\onlinecite{CostaFilho-Almeida-Farias-AndradeJr-2011}.
Likewise, we have that the $\kappa$-deformed Fourier series
(\ref{eq:kappa-Fourier-series})
has the same structure like the proposed by Scarfone
in Ref.~\onlinecite{Scarfone-2015},
considering the $\kappa$-deformed mathematics.
For the particular case $f(x) = 1$, we have
$
f(x) = \lim_{N\rightarrow \infty} f_N(x)
$
with
\begin{equation}
\label{eq:f_N(x)}
f_N(x) = \frac{4}{\pi}\sum_{l=0}^{N}
\frac{1}{2l+1} \sin \left[ (2l+1) \pi
\frac{\textrm{arcsinh} (\kappa x)}{\textrm{arcsinh} (\kappa L)}
\right].
\end{equation}
Similarly as was done in Ref.~\onlinecite{Braga-CostaFilho-2016},
we consider as a quantitative measure of the error the function defined by
$R(N) = \int_0^L [f(x) - f_N(x)]^2 d_\kappa x$.
In Fig.~\ref{fig:4} we show that when $N$ becomes large,
the partial sum $f_N(x)$ converges to $f(x)=1$,
as well as $R(N)$ goes to zero.

%

Expected values of $\hat{x}$ and $\hat{\Pi}_\kappa$
for stationary states can be obtained from usual
internal products of the eigenfunctions $\psi_n (x)$
or, equivalently, from the
deformed internal products of the
modified eigenfunctions  $\varphi (x)$, {\it i.e.},
$
\langle \hat{x}^l \rangle
= \int \psi_n^{\ast} (x) \hat{x}^l \psi_n (x) dx
= \int \varphi_n^{\ast} (x) \hat{x}^l \varphi_n (x) d_\kappa x
$
and
$
\langle \hat{\Pi}_\kappa^l \rangle
= \int \psi_n^{\ast} (x) \hat{\Pi}_\kappa^l  \psi_n (x) dx
= \int \varphi_n^{\ast} (x) \hat{p}_\kappa^l \varphi_n (x) d_\kappa x,
$
which $l$ is a positive integer.
The expectation values
$ \langle \hat{x} \rangle $,
$ \langle \hat{x}^2 \rangle $,
$ \langle \hat{p} \rangle $, and
$ \langle \hat{p}^2 \rangle $
for the eigenstates of the particle in a one dimensional
infinite potential well are respectively
\begin{subequations}
\label{eq:expectation_values_quantum}
\begin{align}
& \label{eq:x-med-quantum}
\frac{\langle \hat{x} \rangle}{L} =
\frac{(\sqrt{1 + \kappa^2 L^2} - 1) (2\pi n)^2 }{
\kappa L \ln \left( \kappa L + \sqrt{1+\kappa^2 L^2} \right)
[\ln^2 \!
\left( \kappa L + \sqrt{1+\kappa^2 L^2} \right) + (2\pi n)^2]},
\\
\label{eq:x^2-med-quantum}
& \frac{\langle \hat{x}^2 \rangle}{L^2} = \frac{1}{2\kappa^2 L^2} \left\{
\frac{\kappa L \sqrt{1+\kappa^2 L^2} (n{\pi})^2}{
\ln \! \left( \kappa L + \sqrt{1+\kappa^2 L^2} \right)
[\ln^2 \! \left( \kappa L + \sqrt{1+\kappa^2 L^2} \right) + (n{\pi})^2]}
- 1 \right\},\\
\label{eq:expected-value-p}
& \langle \hat{p} \rangle = 0,\\
\label{eq:expected-value-p^2}
& \langle \hat{p}^2 \rangle
= \hbar^2 \left[ k_{\kappa,n}^2 \mathcal{I}_{1,0}(1)
+ \kappa^2 \left( \frac{1}{2} \mathcal{I}_{1,0}(1)
-\frac{5}{4}\mathcal{I}_{1,1}(1)
-\mathcal{I}_{3,0}(1)+5\mathcal{I}_{3,1}(1)\right) \right]
\end{align}
\end{subequations}
with
$\mathcal{I}_{j,l}(z)
= 2\int_0^z \textrm{sech}^{2j} (\lambda_\kappa u)
 \textrm{tanh}^{2l} (\lambda_\kappa u) \sin^2 (n\pi u) du$
and
$\lambda_\kappa = \kappa L_\kappa$.
The analytical form of the functions $\mathcal{I}_{j,l}(z)$
is expressed by means of
the Appell hypergeometric function of two variables
(\url{http://functions.wolfram.com/ElementaryFunctions/Sech/21/01/14/01/10/01/0001/})
and due to its complicated expression, it becomes convenient
to write the expectation value (\ref{eq:expected-value-p^2})
in terms of $\mathcal{I}_{j,l}(z) $.

\begin{figure}[!htb]
\centering
 \begin{minipage}[h]{0.32\linewidth}
  \includegraphics[width=\linewidth]{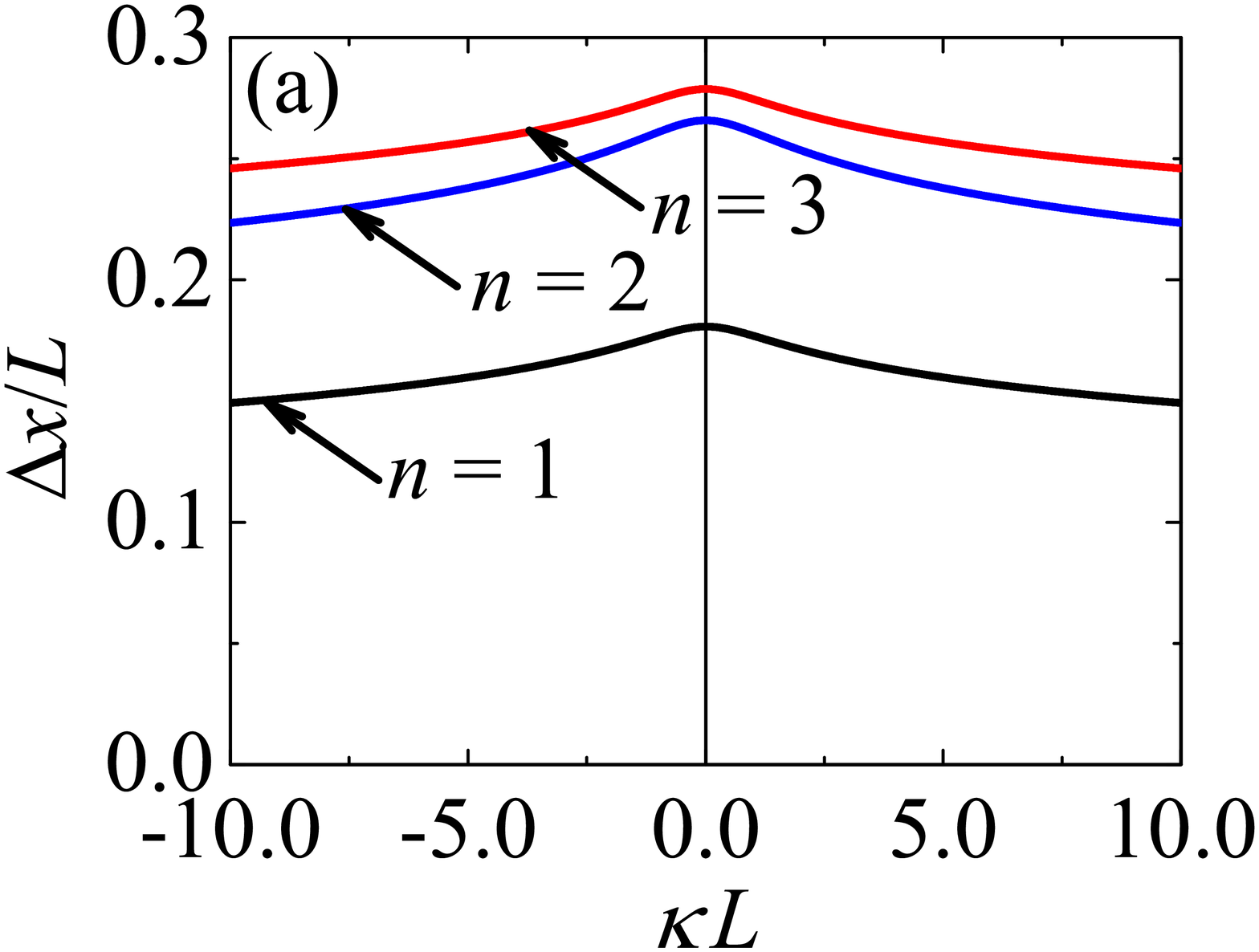}
 \end{minipage}
 \begin{minipage}[h]{0.32\linewidth}
  \includegraphics[width=\linewidth]{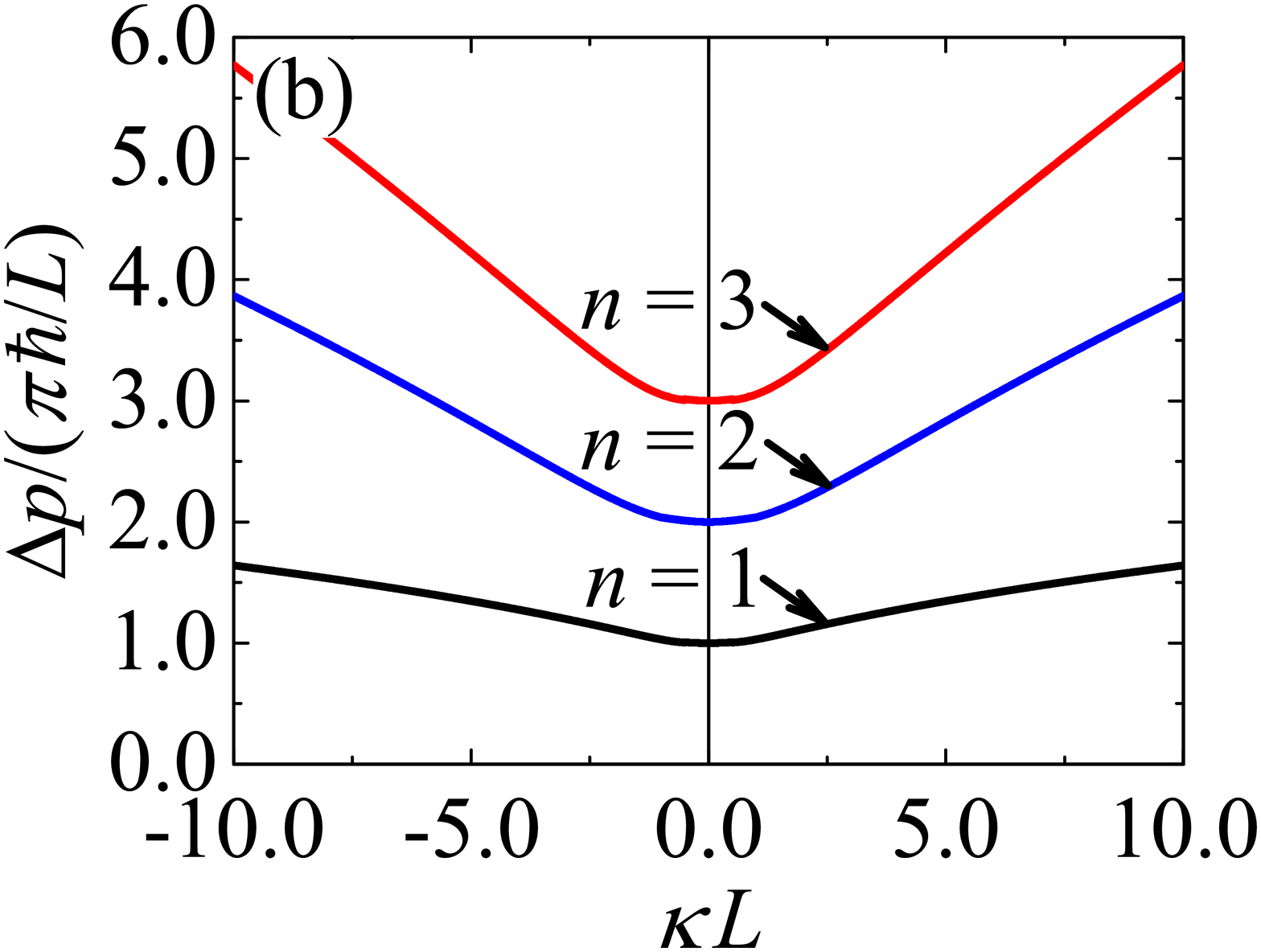}
 \end{minipage}\\
 \begin{minipage}[h]{0.32\linewidth}
  \includegraphics[width=\linewidth]{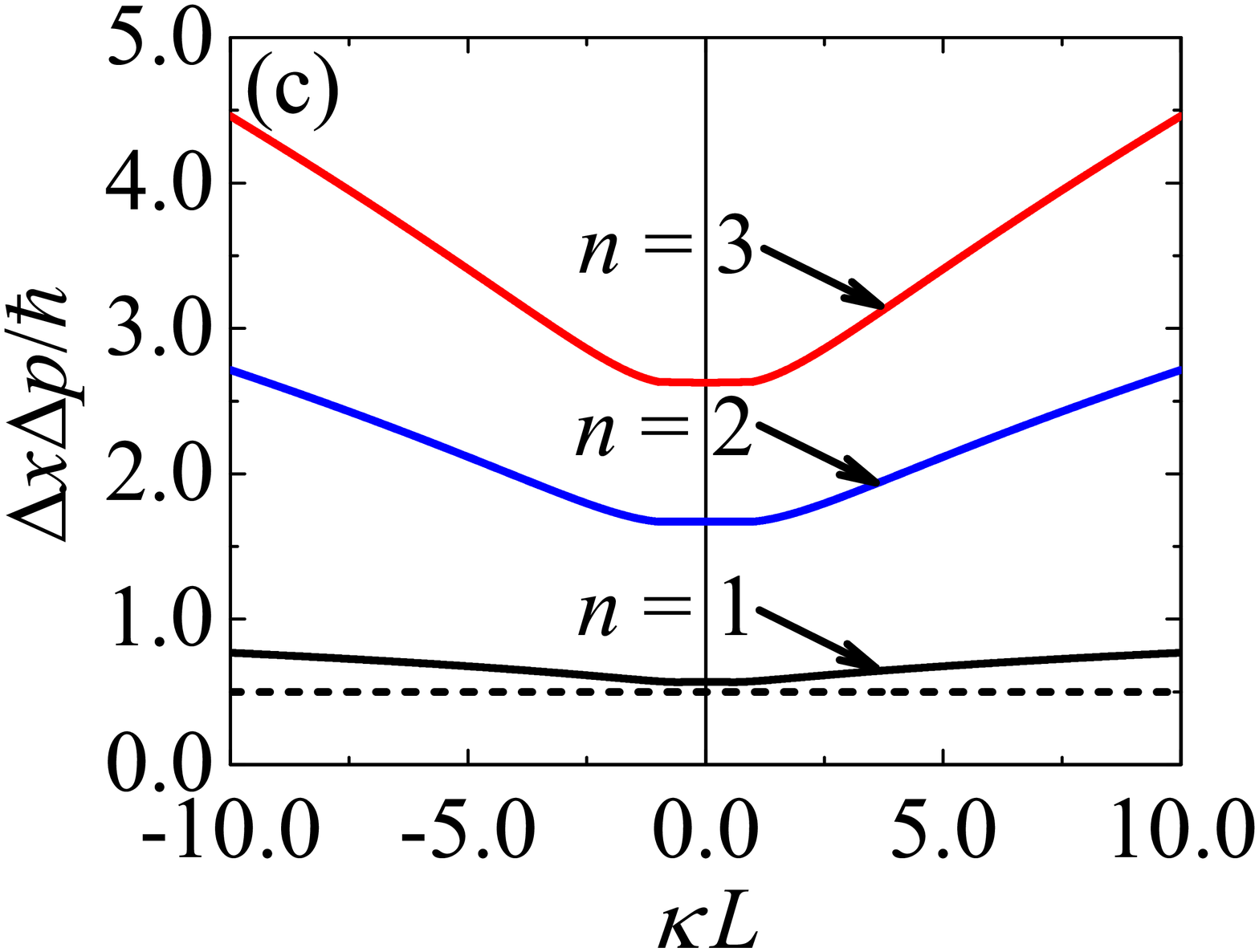}
 \end{minipage}
  \begin{minipage}[h]{0.32\linewidth}
  \includegraphics[width=\linewidth]{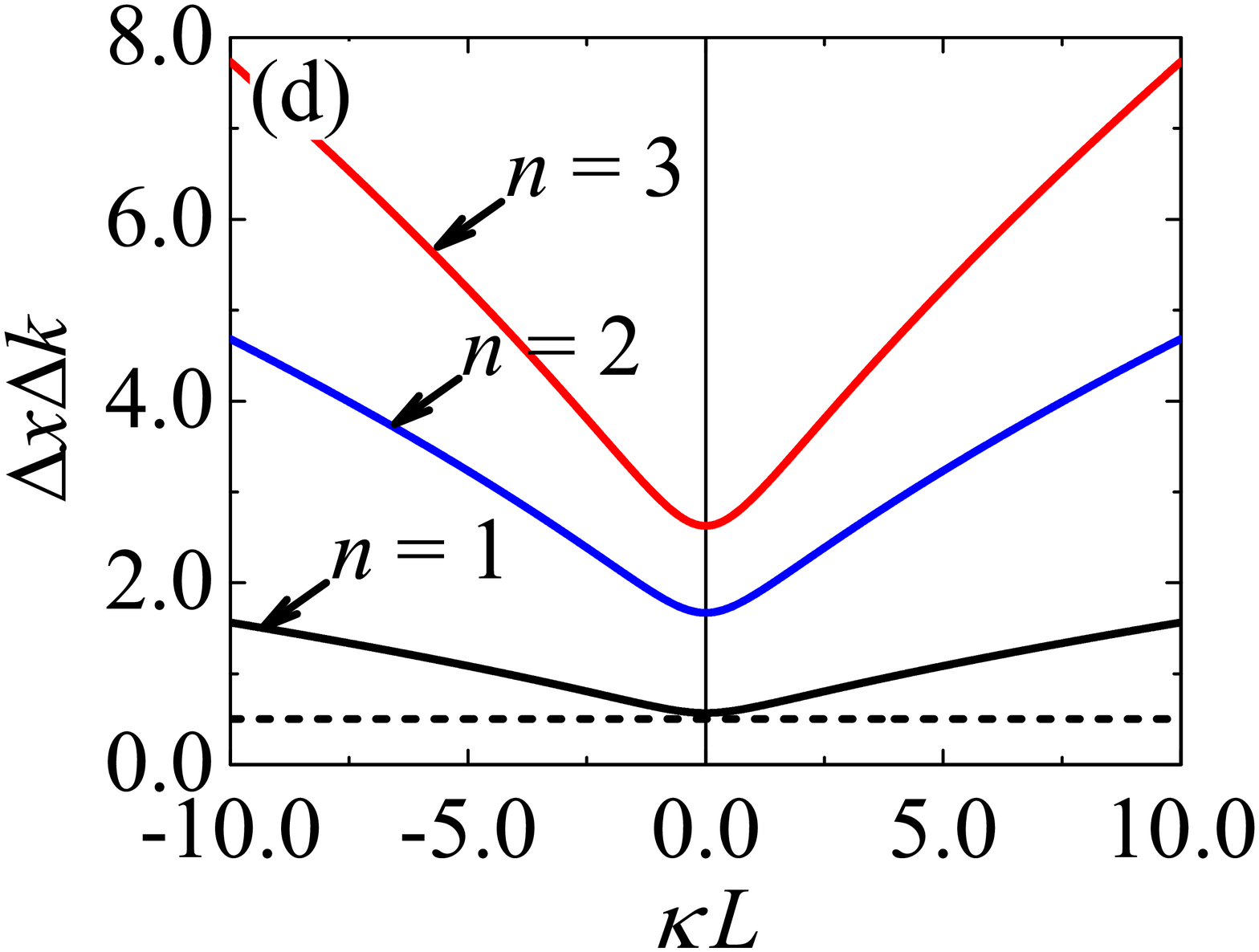}
 \end{minipage}
\caption{\label{fig:5}
(Color online)
Uncertainty in function of $\kappa L$
of (a) the position $\Delta x$, (b) the momentum $\Delta p$
along with the uncertainty relations
((c) and (d)) $\Delta x \Delta p$ and $\Delta x \Delta k$
for a particle with a PDM confined in a box
for the ground state and the first two excited ones.
}
\end{figure}

We can see that in the limit $n \rightarrow \infty$,
the Eqs.~(\ref{eq:expectation_values_quantum})
coincide with the Eqs.~(\ref{eq:classic_first_second_moment}),
which expresses the consistency of the classical limit.
We can also verify that in the limit $\kappa \rightarrow 0$ we
recover the usual results
$ \langle \hat{x} \rangle \rightarrow \frac{L}{2}  $,
$ \langle \hat{x}^2 \rangle \rightarrow \frac{L^2}{3} -
\frac{L^2}{2n^2\pi^2} $ and
$ \langle \hat{p}^2 \rangle \rightarrow \hbar^2k_n^2 $
with
$E_n = \hbar^2 k_n^2 /2m_0$ ($k_n\equiv k_{0,n}=n\pi/L$).
It is straightforwardly to verify that
the expectation values of the pseudo-momentum satisfy
\begin{subequations}
\begin{align}
& \langle \hat{\Pi}_\kappa \rangle =
\hbar \langle k \rangle = 0,\\
& \langle \hat{\Pi}_{\kappa}^2 \rangle =
\hbar \langle k^2 \rangle =
\left( \frac{n\pi\hbar}{L_\kappa} \right)^2
\end{align}
\end{subequations}
with
$\langle \hat{\Pi}_{\kappa}^2 \rangle$ and $\langle \hat{p}^2 \rangle$
different for $\kappa \neq 0$.
In Fig.~\ref{fig:5} we plot the uncertainty relation
for different values of $\kappa$.
Once the operators $\hat{x}$ and $\hat{p}$ are Hermitian and
canonically conjugated, the uncertainty relation is satisfied
for different values of $\kappa$, {\it i.e.},
$\Delta x \Delta p \geq \frac{\hbar}{2}$.
We can also see that position and wave-vector satisfy
the uncertainty relation $\Delta x \Delta k \geq \frac{1}{2}$.
In both curves (c) and (d), the minimum of the uncertainty relation
is attained for $\kappa = 0$.
Similar features have been observed in other system provided with PDM.
In Ref.~\onlinecite{Costa-Gomez-2020} the Cram\'er-Rao, Fisher-Shannon and
L\'opezRuiz-Mancini-Calbet (LMC) complexities have been investigated
for the problem of a particle with a PDM and confined in
an infinite potential well within the framework of the $q$-algebra.
In the context of these complexities, the conjugated variables
exhibit a behavior similar to the standard Heisenberg uncertainty principle.
For different states, the uncertainty relation associated to the
Cram\'er-Rao, Fisher-Shannon and LMC complexities
exhibits a minimum lower bound when the mass of the particle is constant
(i.e., with a null space deformation).
This result is expectedly reasonable since the 
$q$-exponential\cite{Tsallis-Springer-2009} and the $\kappa$-exponential 
functions present a similar behavior when their deformation parameters 
recover the standard exponential ($q \rightarrow 1$ and $\kappa \rightarrow 0$).

\section{\label{sec:kappa-deformed-oscillator}
		$\kappa$-Deformed oscillator with position-dependent mass}

\subsection{$\kappa$-Deformed classical oscillator}

\begin{figure}[!htb]
\centering
\begin{minipage}[h]{0.38\linewidth}
\includegraphics[width=\linewidth]{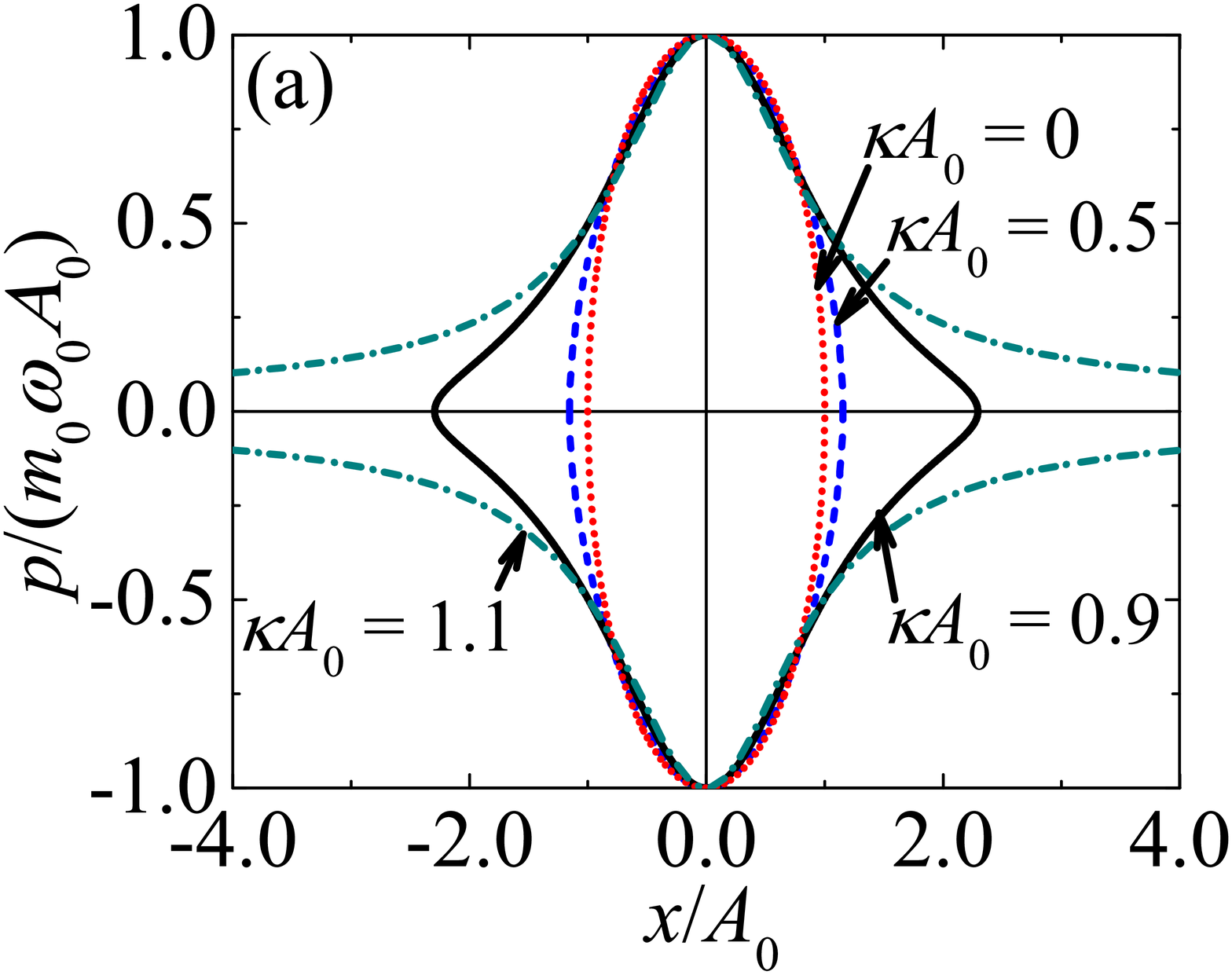}
\end{minipage}
\begin{minipage}[h]{0.38\linewidth}
\includegraphics[width=\linewidth]{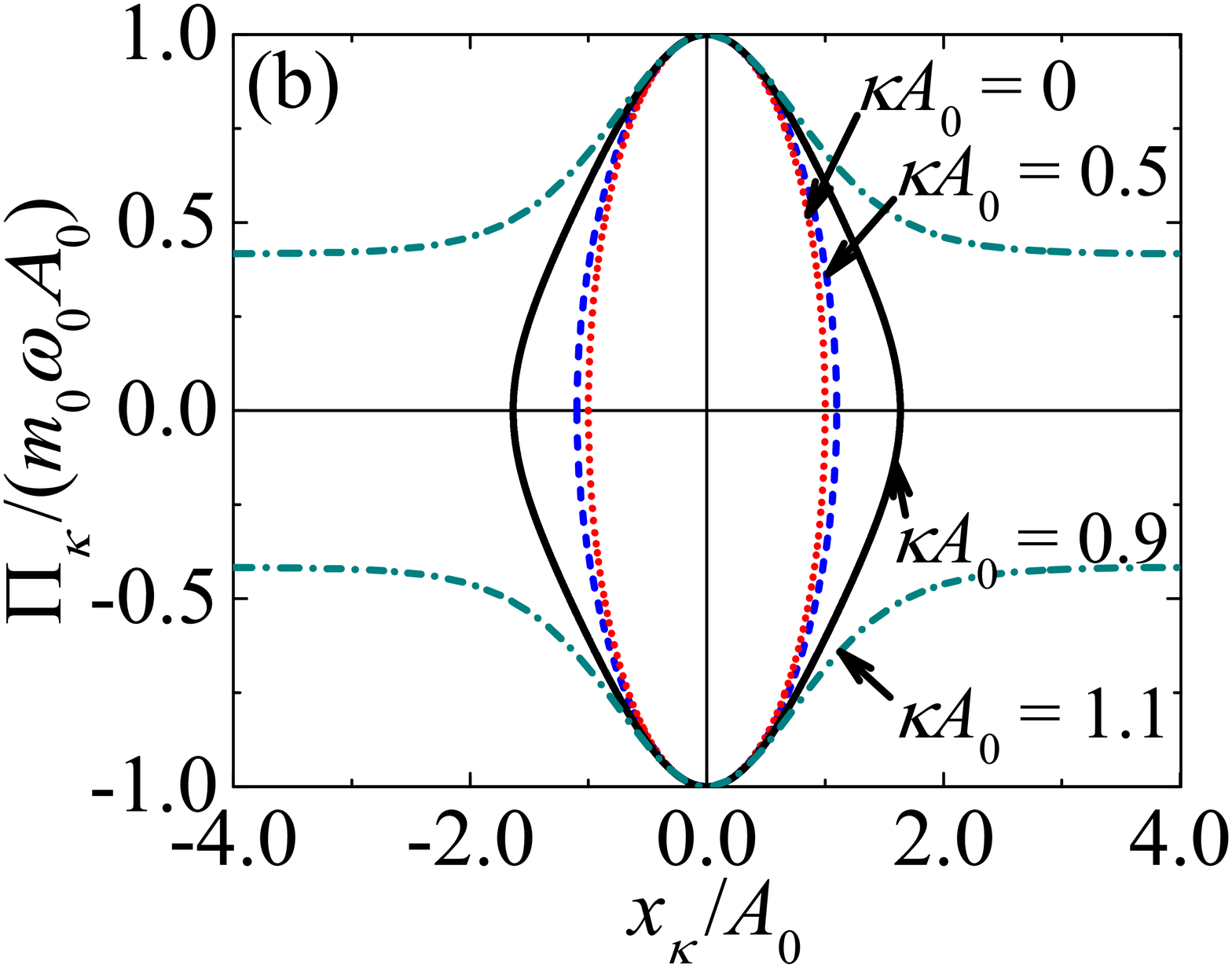}
\end{minipage}
\caption{\label{fig:6}
(Color online)
Phase spaces of the $\kappa$-deformed oscillator in
the (a) usual canonical coordinates $(x, p)$ and
the (b) deformed canonical ones $(x_\kappa, \Pi_\kappa)$
for $\kappa A_0 = 0, 0.5, 0.9$ and $1.1$.
}
\end{figure}

Now we consider a particle with the position-dependent mass (\ref{eq:m(x)})
subjected to the potential $V(x) = \frac{1}{2}m(x)\omega_0^2 x^2$.
This problem is known as the Mathews-Lakshmanan oscillator,
\cite{Mathews-Lakshmanan-1974}
where the classical Hamiltonian is given by
\begin{equation}
\label{eq:H-oscillator}
\mathcal{H}(x,p)=\frac{(1+\kappa^2 x^2)p^2}{2m_0}
				 +\frac{m_0 \omega_0^2 x^2}{2(1+\kappa^2 x^2)}.
\end{equation}
The deformed second Newton's law
(\ref{eq:second_newton_law_generalized})
for this oscillator becomes
\begin{equation}
\label{eq:motion-equation}
\widetilde{D}^2_{\kappa} x (t)
= -\frac{\omega_0^2 x}{(1+\kappa^2 x^2)^2},
\end{equation}
or more explicitly,
\begin{equation}
\label{eq:motion-equation-explicitly}
(1+\kappa^2 x^2) \ddot{x} +\omega_0^2 x
- \kappa^2 \dot{x}^2 x = 0.
\end{equation}
The solution of Eq.~(\ref{eq:motion-equation})
(or equivalently (\ref{eq:motion-equation-explicitly})) is
\begin{equation}
\label{eq:x(t)-osc}
x(t) = A_\kappa \cos ( \Omega_\kappa t + {\delta}_{0}),
\end{equation}
with $A_\kappa = A_0/\sqrt{1-\kappa^2 A_0^2}$
the amplitude of the oscillation,
$\Omega_\kappa = \omega_0 \sqrt{1-\kappa^2 A_0^2}$
the angular frequency and $A_0^2 = 2E/m_0\omega_0^2$.
The potential of this oscillator has a finite well depth
$W_\kappa = m_0\omega_0^2/2\kappa^2$.
Since $E/W_\kappa = \kappa^2 A_0^2$,
the oscillator has a closed (open) path in the phase space
for $0 < \kappa^2 A_0^2 < 1$ ($\kappa^2 A_0^2 > 1$),
according to Ref.~\onlinecite{Mathews-Lakshmanan-1974}.
The PCT (\ref{eq:x_k-Pi_k})
maps the Hamiltonian (\ref{eq:H-oscillator})
into the corresponding to the anharmonic oscillator,
{\it i.e.}
\begin{equation}
\mathcal{K}(x_{\kappa}, \Pi_{\kappa})
= \frac{1}{2m_0}\Pi_\kappa^2
+ W_\kappa \tanh^2 (\kappa x_{\kappa}),
\end{equation}
with $\kappa$ a continuous parameter that controls
the anharmonicity of the potential.
In Fig.~\ref{fig:6} we plot the phase spaces $(x, p)$ and
$(x_\kappa, \Pi_\kappa)$ for different values of $\kappa A_0$.
The bounded motion in the interval
$-A_\kappa < x < A_\kappa$ of
the
standard space turns out into the interval
$-x_{\kappa,\textrm{max}} < x_\kappa <
x_{\kappa,\textrm{max}}
=\kappa^{-1}\textrm{atanh}(\kappa A_0)$
in the deformed space.
Besides, the unbounded motion has the interval of
the linear momentum $0 < |p| < m_0 \omega_0 A_0$
turned into
$m_0 \omega_0 A_0 \sqrt{1-\frac{1}{\kappa^2 A_0^2}}
< |\Pi_\kappa| < m_0 \omega_0 A_0$.
As the dimensionless parameter $\kappa A_0$ increases from $0$ to $1.1$
within the interval $[0.9,1.1]$ it is observed that the horizontal axe
of the ellipses become infinite, thus giving place to an unbounded motion.

By means of the WKB approximation we can obtain the energy levels of
the corresponding quantum system.
Using this method, we have
\begin{align}
\left( n + \frac{1}{2} \right) \frac{\hbar}{2} &=
\frac{1}{2\pi} \int_{-A_\kappa}^{A_\kappa} p(x) dx
= \frac{m_0 \Omega_\kappa}{2\pi}
\int_{-A_\kappa}^{A_\kappa}
\frac{\sqrt{A_{\kappa}^2-x^2}}{1+\kappa^2 x^2} dx
\nonumber \\
&=  \frac{m_0 \Omega_{\kappa} A_{\kappa}^2 }{4\pi}
\int_{0}^{2\pi} \frac{\sin^2 \theta_\kappa}{
1+\kappa^2 A_{\kappa}^2 \cos^2 \theta_\kappa}
d\theta_\kappa
\nonumber \\
&=  \frac{m_0 \Omega_\kappa}{2{\kappa}^2}
\left( \sqrt{1+{\kappa}^2 A_{\kappa}^2} -1 \right)
\end{align}
with $n=0,1,2,\ldots$. Since
$E=  \frac{1}{2}m_0 \Omega_{\kappa}^2 A_\kappa^2$
we obtain
\begin{equation}
\label{eq:E_n_by-WKB}
E_n = \hbar \omega_0 \left( n + \frac{1}{2} \right)
	 -\frac{\hbar^2 \kappa^2}{2m_0}
	 \left( n + \frac{1}{2} \right)^2,
\end{equation}
which corresponds to the energy levels of an
anharmonic oscillator.

From Eq.~(\ref{eq:x(t)-osc}), the classical density probability
of finding the particle between $x$ and $x + dx$ results
$
\rho_{\textrm{classic}}(x) = \frac{1}{\pi\sqrt{A_\kappa^2 - x^2}}.
$
The first and second moments of the position and the linear momentum
in terms of the amplitude or the energy
for the deformed oscillator are
\begin{subequations}
\label{eq:x-p-x^2-p^2-expec-values}
\begin{align}
\label{eq:x-med-osc}
& \overline{x}  = 0,\\
\label{eq:x^2-med-osc}
& \overline{x^2} =  \frac{A_\kappa^2}{2} = \frac{E}{
m_0 \omega_0^2 \left( 1-\frac{2E{\kappa}^2}{m_{0}{\omega}_0^2}\right)},\\
\label{eq:p-med-osc}
& \overline{p} = 0,\\
\label{eq:p^2-med-osc}
& \overline{p^2} =
\frac{m_0 \omega_0^2 A_\kappa^2}{2(1+\kappa^2 A_\kappa^2)^{3/2}}
 = m_0 E  \sqrt{1-\frac{2E{\kappa}^2}{m_{0}{\omega}_0^2}}.
\end{align}
\end{subequations}
The mean values of the kinetic and potential energies satisfy
the relationship
\begin{equation}
\label{eq:T-average-value}
\overline{T} = E - \overline{V}
= \frac{m_0 \omega_0^2}{2\kappa^2}
\frac{1}{\sqrt{1+\kappa^2 A_{\kappa}^2}}
\left(
1-\frac{1}{\sqrt{1+\kappa^2 A_{\kappa}^2}}
\right),
\end{equation}
with $\overline{V} = \int \rho_{\textrm{classic}}(x) V(x) dx$.
Since $\overline{V}=\overline{T}/\sqrt{1-\kappa^2 A_0^2} $,
we have that the virial theorem ($\overline{V} = \overline{T}$)
is satisfied only for $\kappa A_0=0$, which implies $\kappa=0$.

\subsection{$\kappa$-Deformed quantum oscillator}

\begin{figure}[!htb]
\centering
\begin{minipage}[h]{0.38\linewidth}
\includegraphics[width=\linewidth]{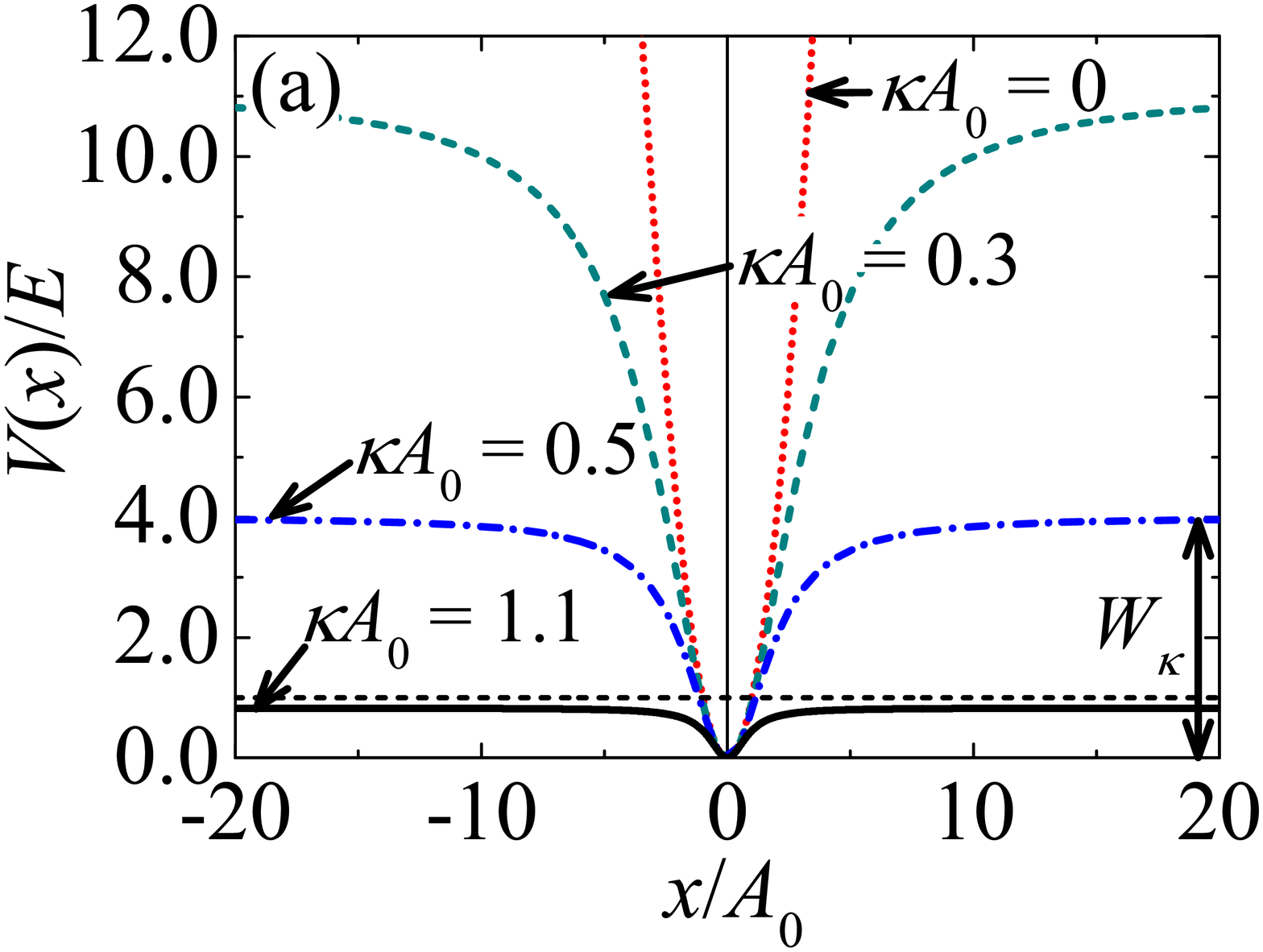}
\end{minipage}
\begin{minipage}[h]{0.38\linewidth}
\includegraphics[width=\linewidth]{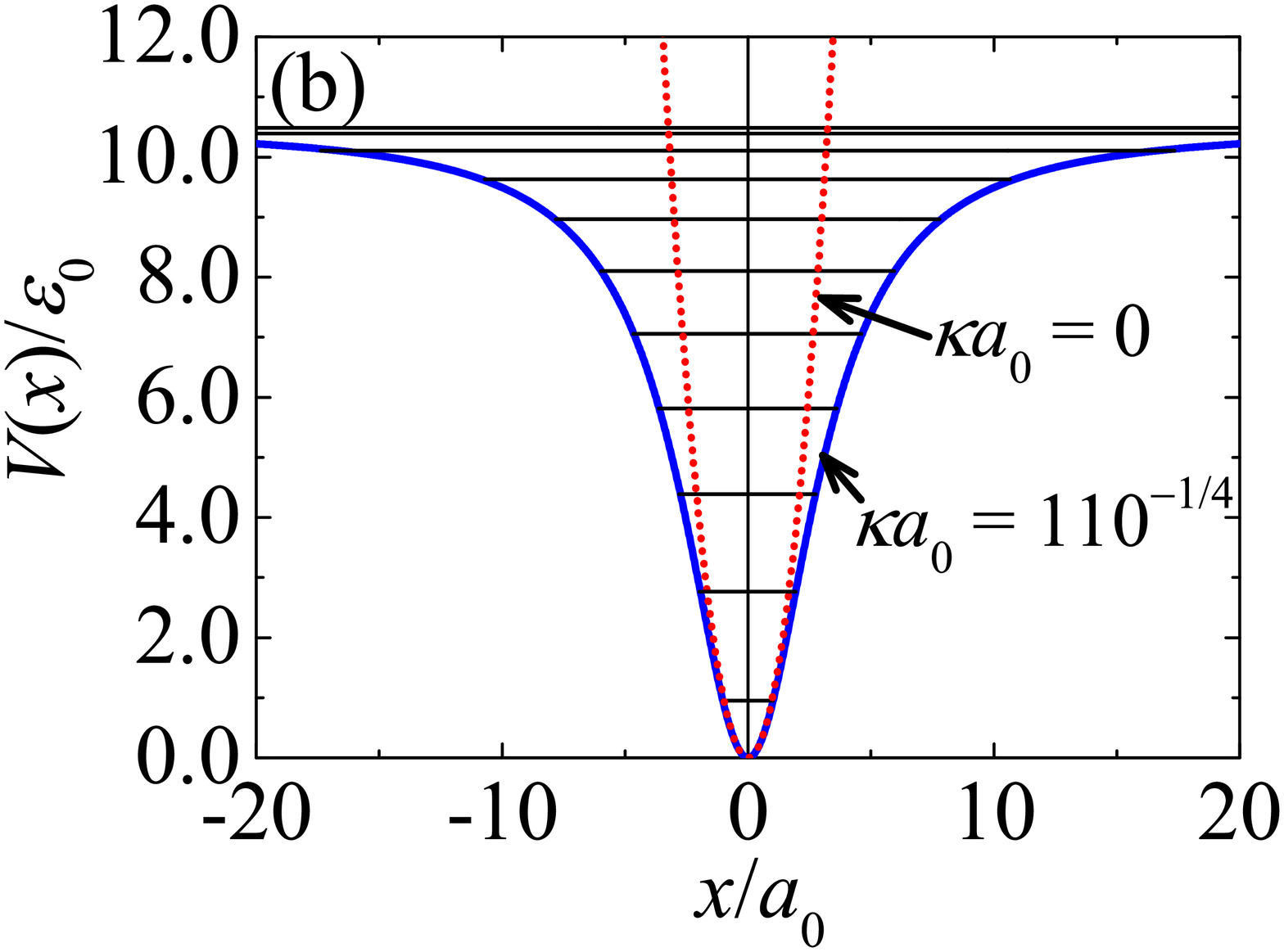}
\end{minipage}
\caption{\label{fig:7}
(Color online)
(a) Potential $V(x) =
\frac{m_0 \omega_0^2 x^2}{2(1+\kappa^2 x^2)}$
for $\kappa A_0 = 0, 0.3, 0.5$ and $1.1$
with $A_0^2 = 2E/m_0\omega_0^2$.
(b) Energy levels of the $\kappa$-deformed oscillator for
$\kappa a_0 = 1/\sqrt[4]{110}$ with $a_0 = \hbar/m_0 \omega_0$
and $\varepsilon_0 = \hbar \omega_0/2$.
}
\end{figure}

\begin{figure}[!htb]
\centering
\begin{minipage}[h]{0.32\linewidth}
\includegraphics[width=\linewidth]{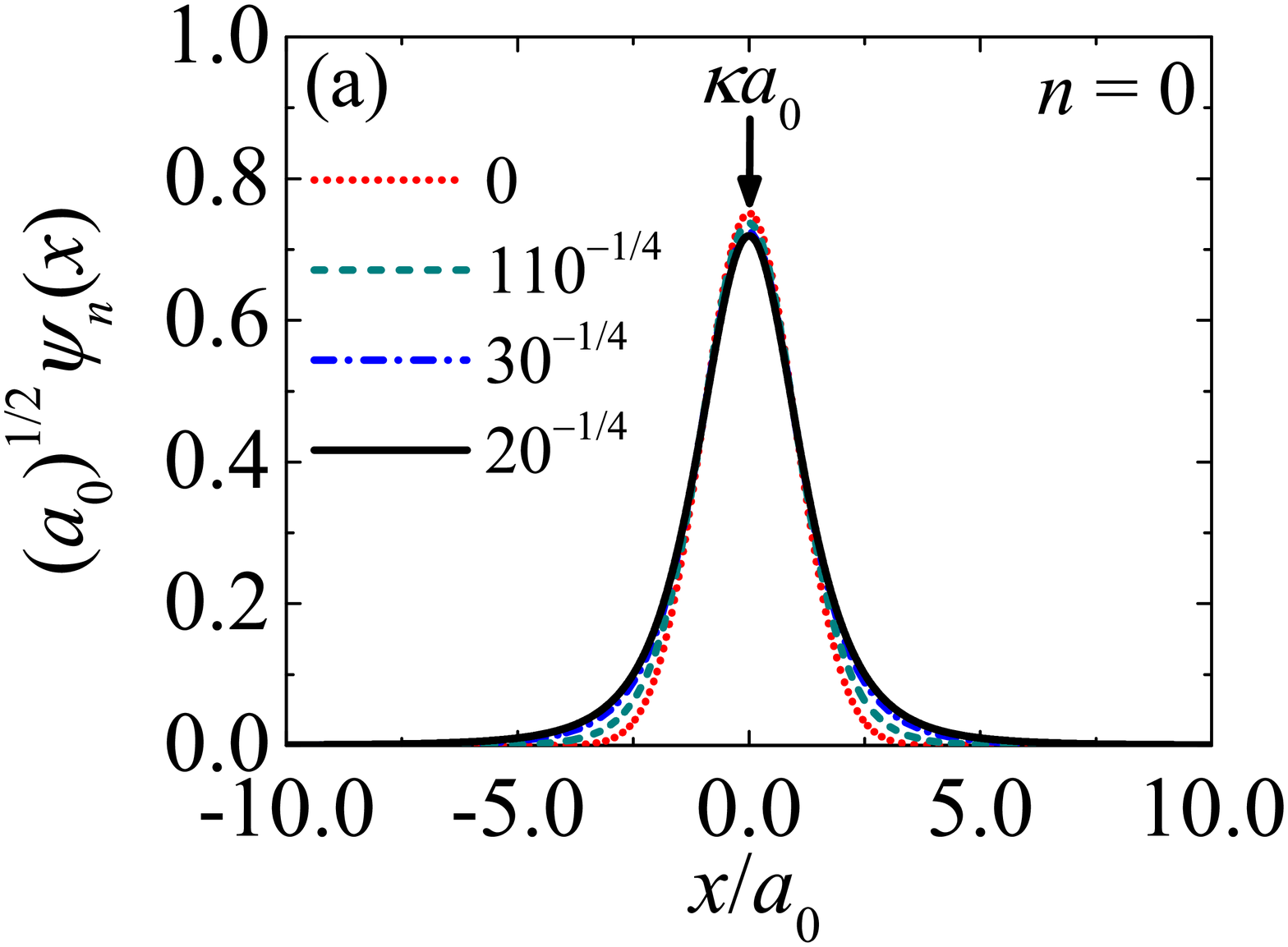}
\end{minipage}
\begin{minipage}[h]{0.32\linewidth}
\includegraphics[width=\linewidth]{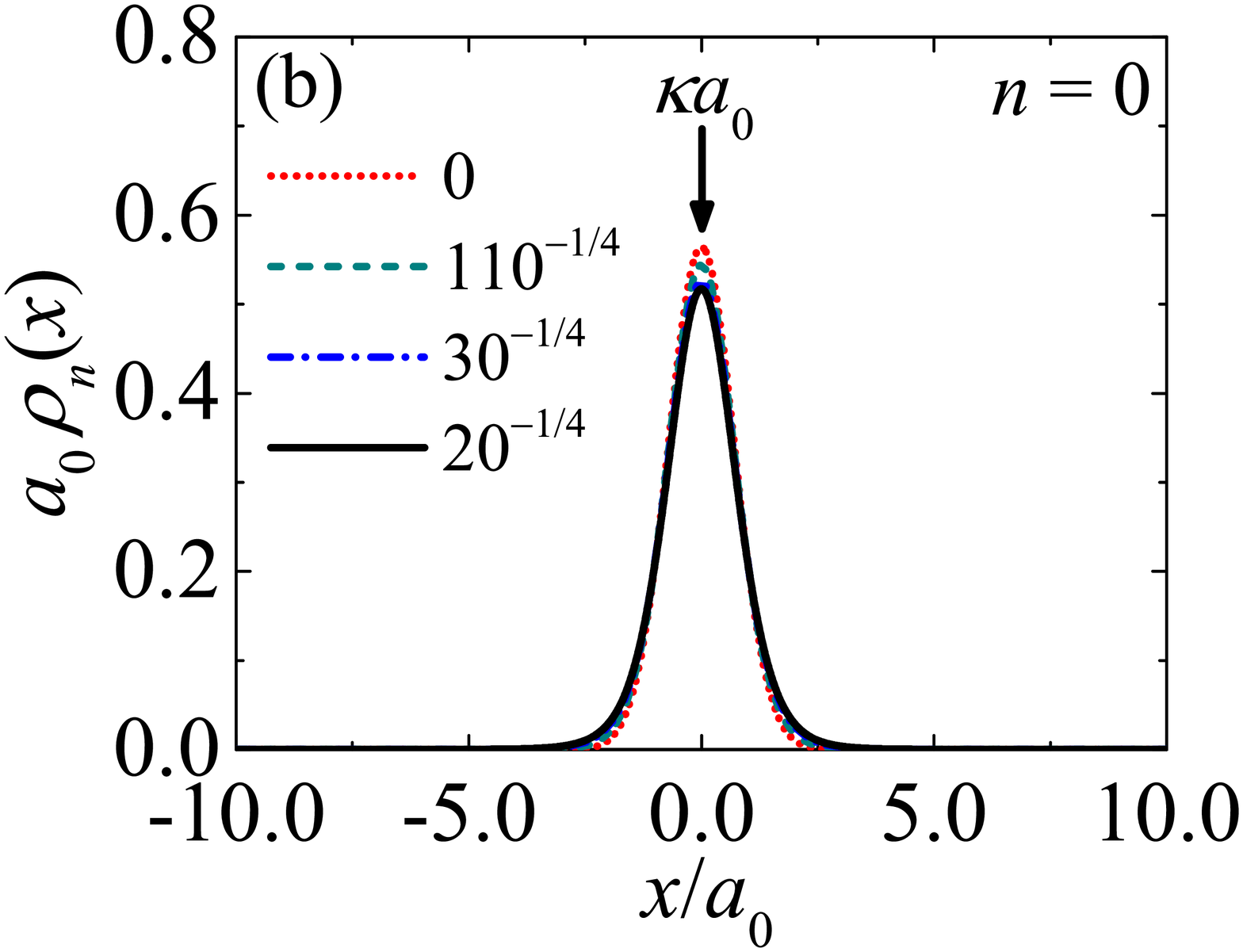}
\end{minipage}\\
\begin{minipage}[h]{0.32\linewidth}
\includegraphics[width=\linewidth]{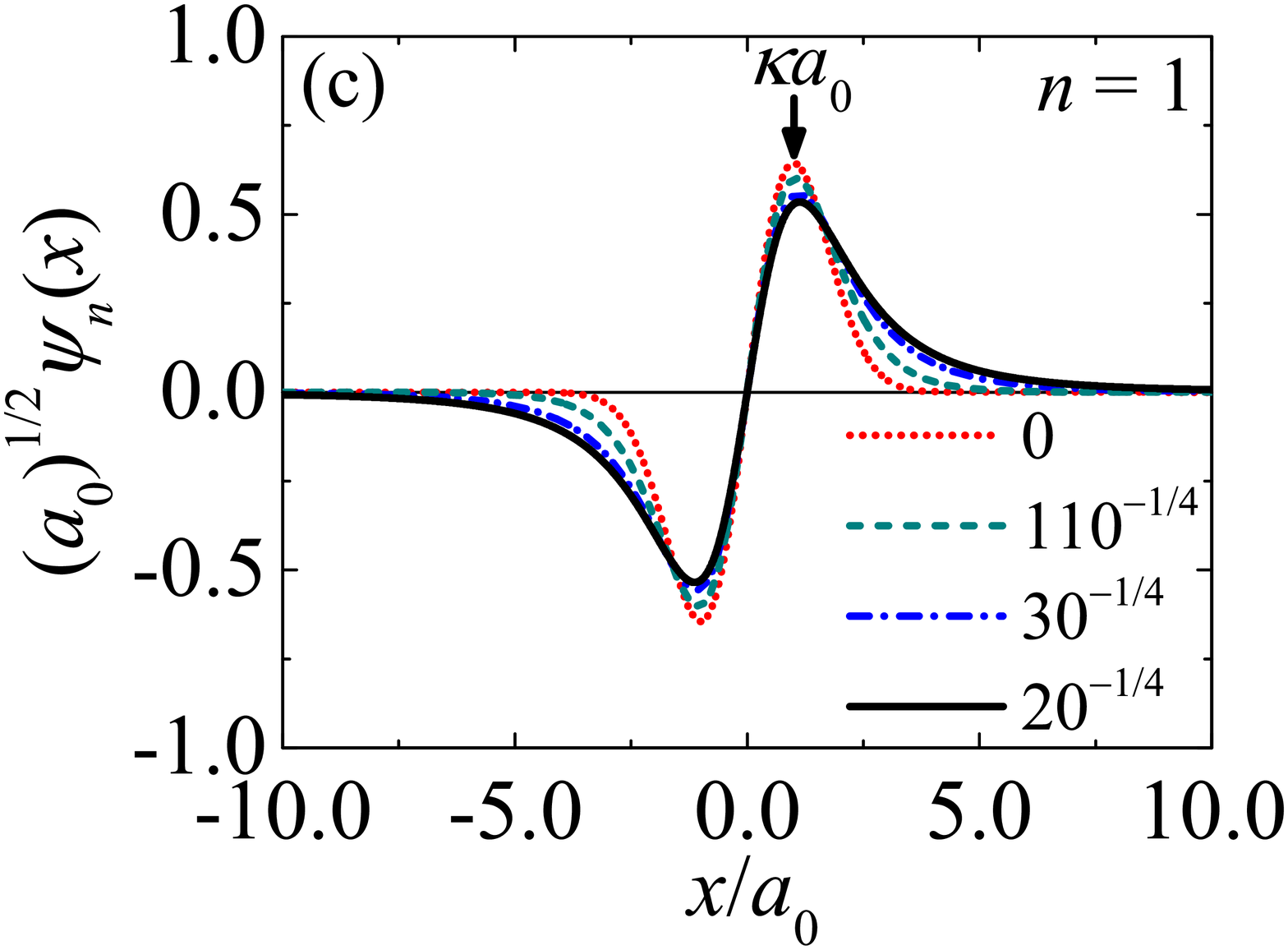}
\end{minipage}
\begin{minipage}[h]{0.32\linewidth}
\includegraphics[width=\linewidth]{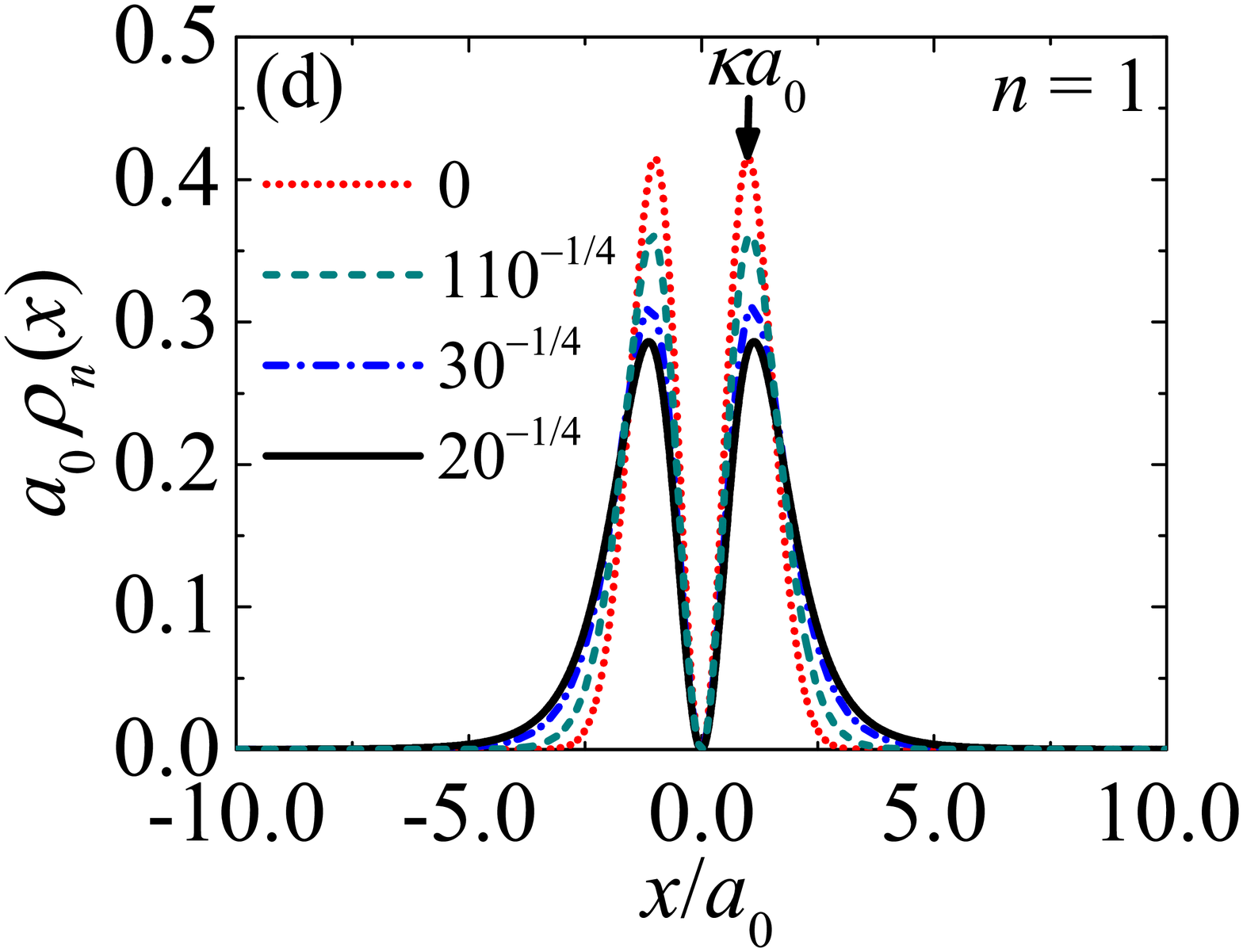}
\end{minipage} \\
\begin{minipage}[h]{0.32\linewidth}
\includegraphics[width=\linewidth]{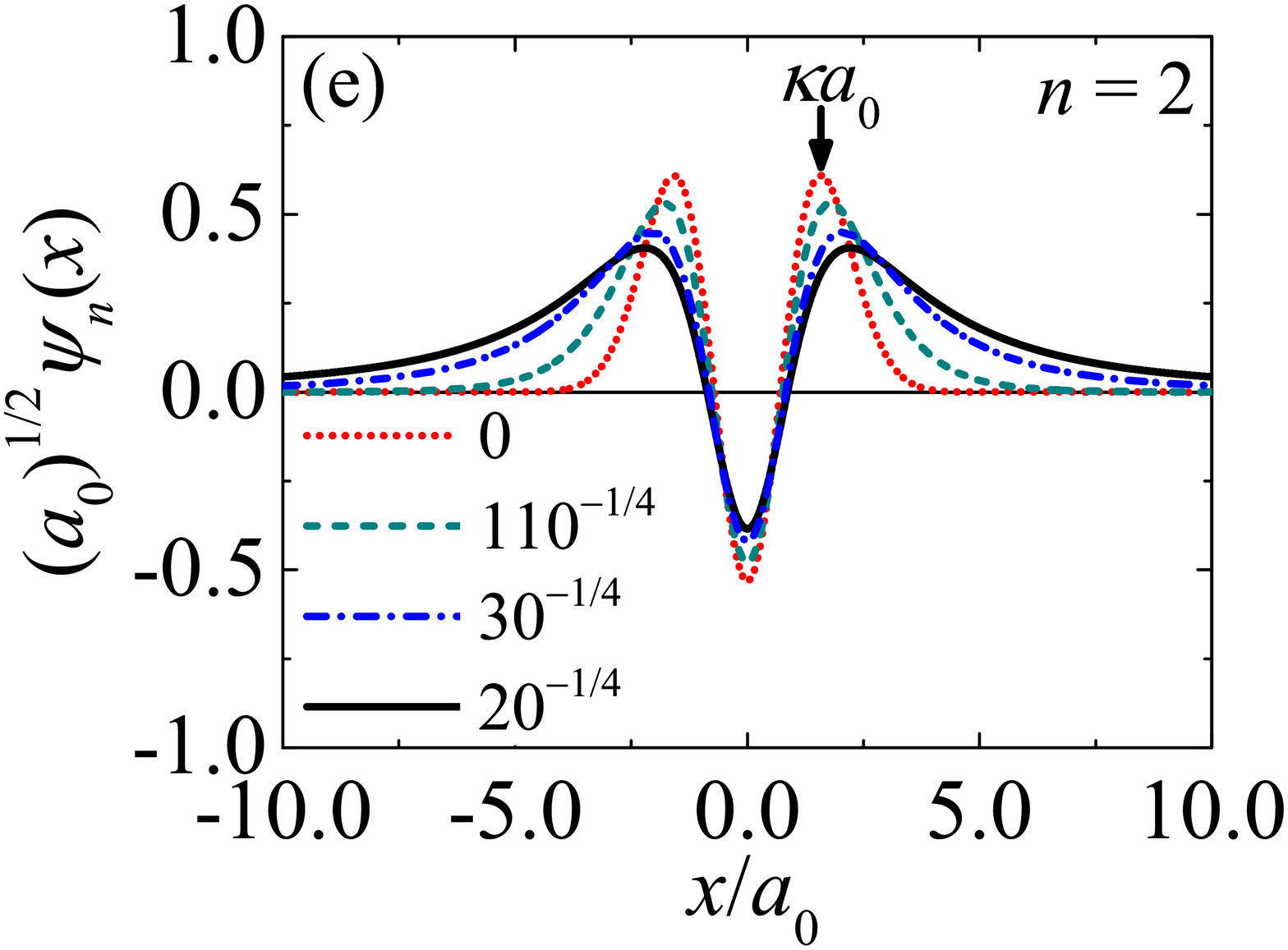}
\end{minipage}
\begin{minipage}[h]{0.32\linewidth}
\includegraphics[width=\linewidth]{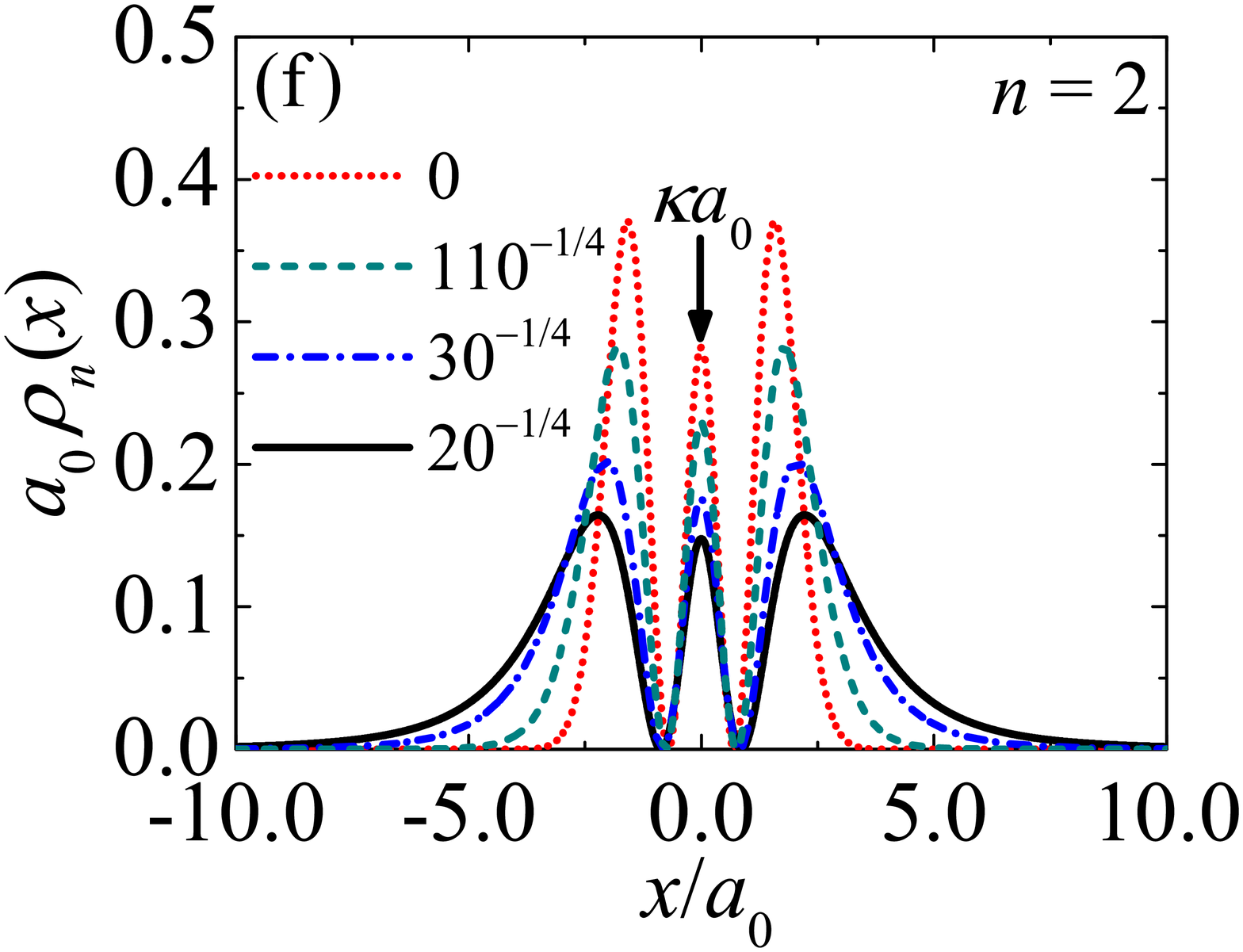}
\end{minipage}\\
\begin{minipage}[h]{0.32\linewidth}
\includegraphics[width=\linewidth]{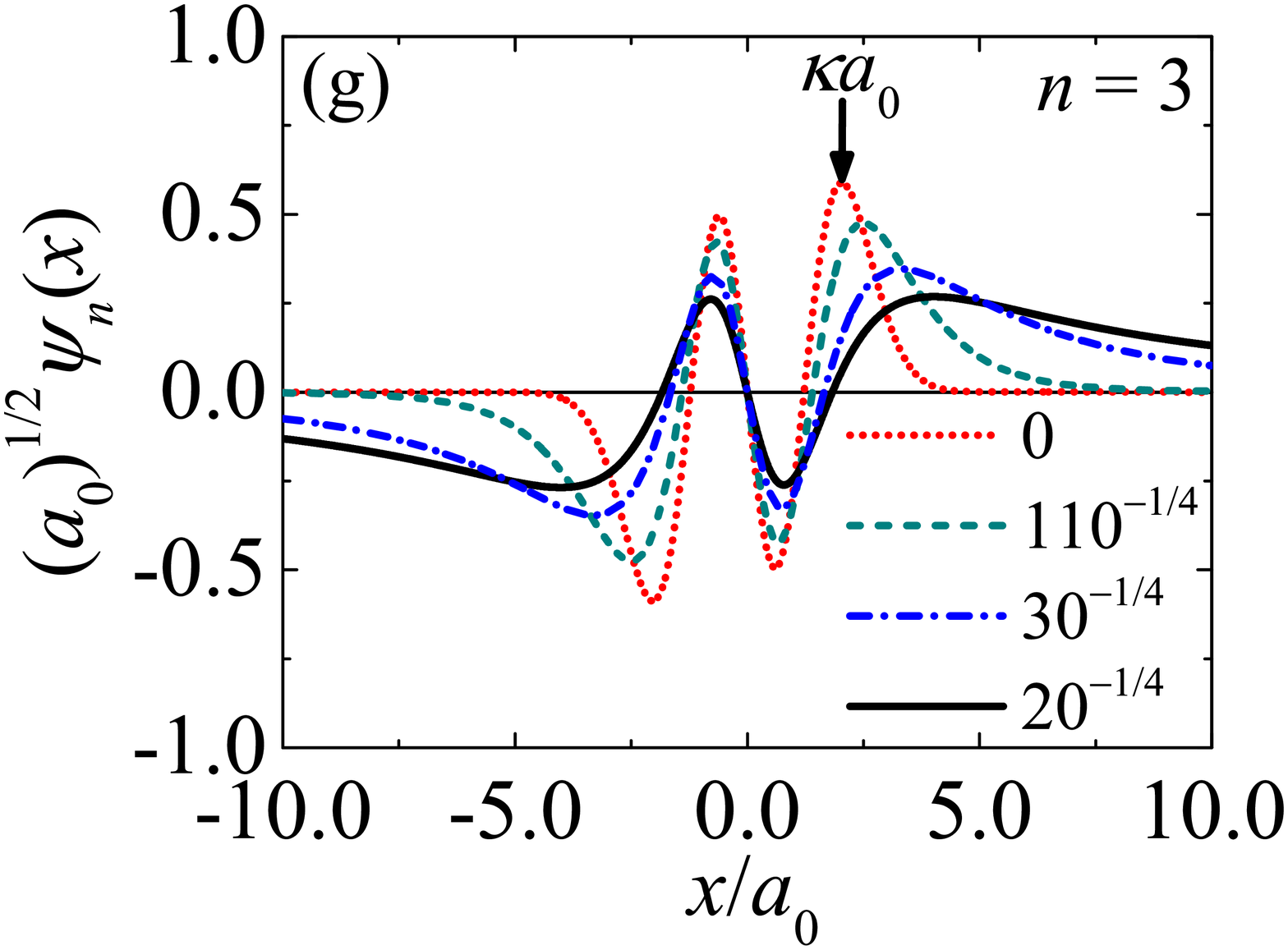}
\end{minipage}
\begin{minipage}[h]{0.32\linewidth}
\includegraphics[width=\linewidth]{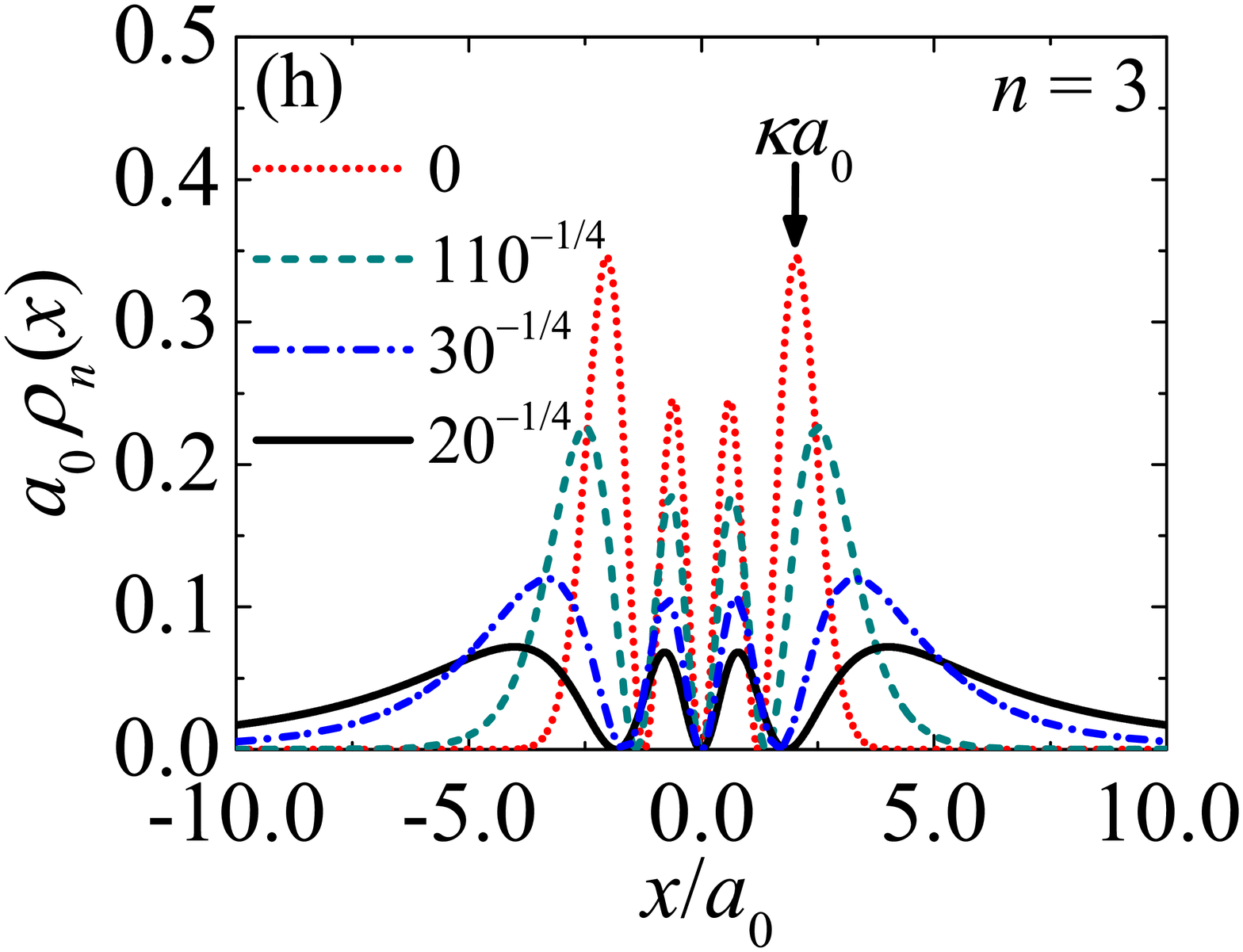}
\end{minipage}\\
\caption{\label{fig:8}
(Color online)
Eigenfunctions $\psi_n (x)$ (upper line)
and probability densities $\rho_n(x) = |\psi_n (x)|^2 $ (bottom line)
for a $\kappa$-deformed oscillator particle
for the values of $\kappa a_0$ such that
$\nu(\nu+1)=1/(\kappa a_0)^4$ with
$\nu = 4, 5, 10$ and $\infty$
in such a way that the corresponding values of
$\kappa a_0$ are $20^{-1/4},30^{-1/4}, 110^{-1/4}$ and $0$.
(a) and (b): $n=0$ (ground state),
(c) and (d): $n=1$ (first excited state),
(e) and (f): $n=2$ (second excited state),
(g) and (h): $n=3$ (third excited state).
}
\end{figure}

The corresponding $\kappa$-deformed time-independent Schr\"odinger
equation for the PDM oscillator is
\cite{Mathews-Lakshmanan-1975}
\begin{equation}
\label{eq:deformed-schrodinger-equation-osc}
	-\frac{\hbar^2}{2m_0} D_{\kappa}^2 \varphi (x)
	+ \frac{1}{2}\frac{m_0 \omega_0^2 x^2}{(1+\kappa^2 x^2)}
    \varphi (x)= E\varphi (x).
\end{equation}
Making the change of variable
$x\rightarrow x_\kappa
= \kappa^{-1}\textrm{arcsinh} (\kappa x)$
(see Eq.~(\ref{eq:transformation-u_kappa-u}))
we obtain a particle with constant mass $m_0$
subjected to the P\"oschl-Teller potential
\begin{equation}\label{eq:Poschl-Teller}
 -\displaystyle \frac{{\hbar}^2}{2m_0} \frac{d^2\phi(x_\kappa)}{dx_\kappa^2}
  -\frac{\hbar{^2 {\kappa}^2}}{m_0} \frac{\nu ({\nu}+1)}{2}
 \textrm{sech}^2 (\kappa x_{\kappa}) \phi (x_{\kappa})
 = \epsilon \phi (x_{\kappa}),
\end{equation}
with
$\epsilon = E-\hbar \omega_0/2\kappa^2 a_0^2$,
$\nu(\nu+1)=1/\kappa^4 a_0^4$ and
$a_0^2 = \hbar/m_0 \omega_0$.
The solutions of the
Eq.~(\ref{eq:Poschl-Teller})
are
\begin{equation}
\phi (x_{\kappa}) =
\sqrt{\frac{\kappa \mu ({\nu - \mu})!}{({\nu}+{\mu})!}}
P_\nu^{\mu} (\textrm{tanh}(\kappa x_{\kappa})),
\end{equation}
where $\mu = \nu - n$, $n$ is an integer
and $P_\nu^{\mu}$ are the associated Legendre polynomials.
Then, the eigenfunctions for
the $\kappa$-deformed oscillator in the space representation $x$ are
\begin{align}
\psi_{n} (x) &=
\sqrt{\frac{\kappa (\nu - n) n!}{({2\nu}-n)!}}
\frac{1}{\sqrt[4]{1+\kappa^2 x^2}}
P_\nu^{\nu - n} \left( \frac{\kappa x}{\sqrt{1+{\kappa}^2 x^2}} \right).
\end{align}
The energy levels are given by
\begin{equation}
\label{eq:energy-levels}
E_n = \hbar \omega_\kappa \left( n+\frac{1}{2} \right)
	  -\frac{\hbar^2 \kappa^2}{2m_0} \left( n+\frac{1}{2} \right)^2
	  -\frac{\hbar^2 \kappa^2}{8m_0}
\end{equation}
with $\omega_\kappa = \omega_0 \sqrt{1+\frac{\hbar^2 \kappa^4 }{4m_0^2 \omega_0^2}}$.
It should be noted that the quantum energy levels
differ from those obtained using
the WKB approximation (Eq.~(\ref{eq:E_n_by-WKB}))
by the constant term $-\frac{\hbar^2 \kappa^2}{8m_0}$
and the frequency of small oscillations $\omega_0$
replaced by $\omega_\kappa$.
This modification in the frequency is associated with the
symmetrization problem of the classical
Hamiltonian in order to construct
its corresponding Hamiltonian operator
in the quantum formalism
(see Ref.~\onlinecite{Mathews-Lakshmanan-1975} for more details).
However, in the limit $\hbar \rightarrow 0$
with $n \gg 1$, the Eq.~(\ref{eq:energy-levels})
recovers the semi-classical approximation,
Eq.~(\ref{eq:E_n_by-WKB}).
In Fig.~\ref{fig:7} an illustration of the potential
$V(x)=\frac{m_0\omega_0^2x^2}{2(1+\kappa^2 x^2)}$
along with the energy levels for some values of $\kappa A_0$, is shown.
In Fig.~\ref{fig:8} we show the wave-functions and
the probability densities for the four lower energy states and
for some values of $\kappa a_0$.
The values of $\kappa a_0$ chosen are such that
$\nu(\nu+1)=1/\kappa^4 a_0^4$ is satisfied with $\nu$ integer.
We consider $\nu = 4, 5, 10$ and $\infty$
in such a way that the corresponding values of
$\kappa a_0$ are $20^{-1/4},30^{-1/4}, 110^{-1/4}$ and $0$.

From the Legendre differential equation
\begin{equation}
(1-u^2) \frac{d^2 P_\nu^{\mu}(u)}{du^2}
- 2u \frac{d P_\nu^{\mu}(u)}{du}
+ \left[\nu (\nu + 1) - \frac{\mu^2}{1-u^2}
\right] P_\nu^{\mu}(u) = 0,
\end{equation}
the identities (see Eqs.~(2) and (3) in page 965
of the Ref.~\onlinecite{Gradshteyn}),
we obtain the expectation values of $ \langle \hat{x} \rangle $,
$ \langle \hat{x}^2 \rangle $,
$ \langle \hat{p} \rangle $, and
$ \langle \hat{p}^2 \rangle $ are
\begin{subequations}
\label{eq:x-p-x^2-p^2-expec-values-E_n}
\begin{eqnarray}
\label{eq:x-quantum}
\langle \hat{x} \rangle &=& 0,
\\
\label{eq:x^2-quantum}
\langle \hat{x}^2 \rangle &=&
\frac{E_n + \frac{\hbar^2 \kappa^2}{2m_0}}{
m_0\omega_0^2 \left( 1 - \frac{2E_n \kappa^2}{m_0 \omega_0^2} -
\frac{\hbar^2 \kappa^4}{m_0^2 \omega_0^2}\right)}
\nonumber \\
& = & \frac{\hbar}{m_0 \omega_0}
	\left\{
	\frac{\frac{\omega_\kappa}{\omega_0} \left( n + \frac{1}{2} \right)
	-\frac{\kappa^2 a_0^2}{2} \left( n^2 + n -\frac{1}{2}\right)}{
	1 - 2\kappa^2 a_0^2 \left[
	\frac{\omega_\kappa}{\omega_0} \left( n + \frac{1}{2} \right)
	-\frac{\kappa^2 a_0^2}{2} \left( n^2 + n -\frac{1}{2}\right)
	\right]} \right\},
\\
\label{eq:p-quantum}
\langle \hat{p} \rangle &=& 0,
\\
\label{eq:p^2-quantum}
\langle \hat{p}^2 \rangle &=&
m_0 \left( E_n - \frac{\hbar^2 \kappa^2}{4m_0} \right)
\frac{z^2 - (2n+1)z}{z^2-4}
\nonumber \\
&=& m_0 \hbar \omega_0 \left[
	\frac{\omega_\kappa}{\omega_0}
	\left( n + \frac{1}{2} \right)
	-\frac{\kappa^2 a_0^2}{2} (n^2 + n + 1)
	\right] \frac{z^2 - (2n+1)z}{z^2-4},
\end{eqnarray}
\end{subequations}
with $z \equiv 2\nu + 1
= \sqrt{1+\frac{4m_0^2 \omega_0^2}{\kappa^4 \hbar^2}}$.

\begin{figure*}[!hbt]
\centering
\begin{minipage}[h]{0.32\linewidth}
\includegraphics[width=\linewidth]{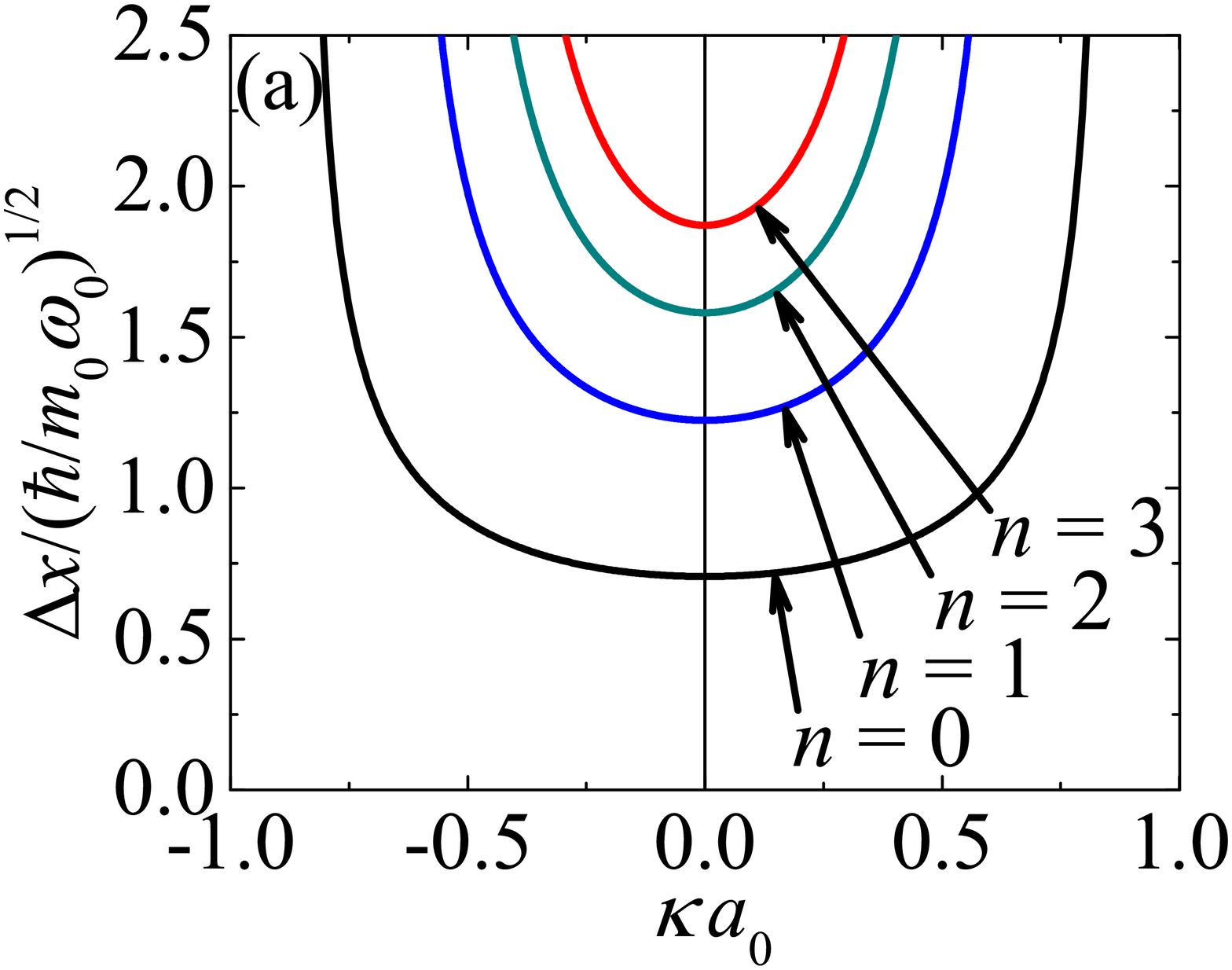}
\end{minipage}
\begin{minipage}[h]{0.32\linewidth}
 \includegraphics[width=\linewidth]{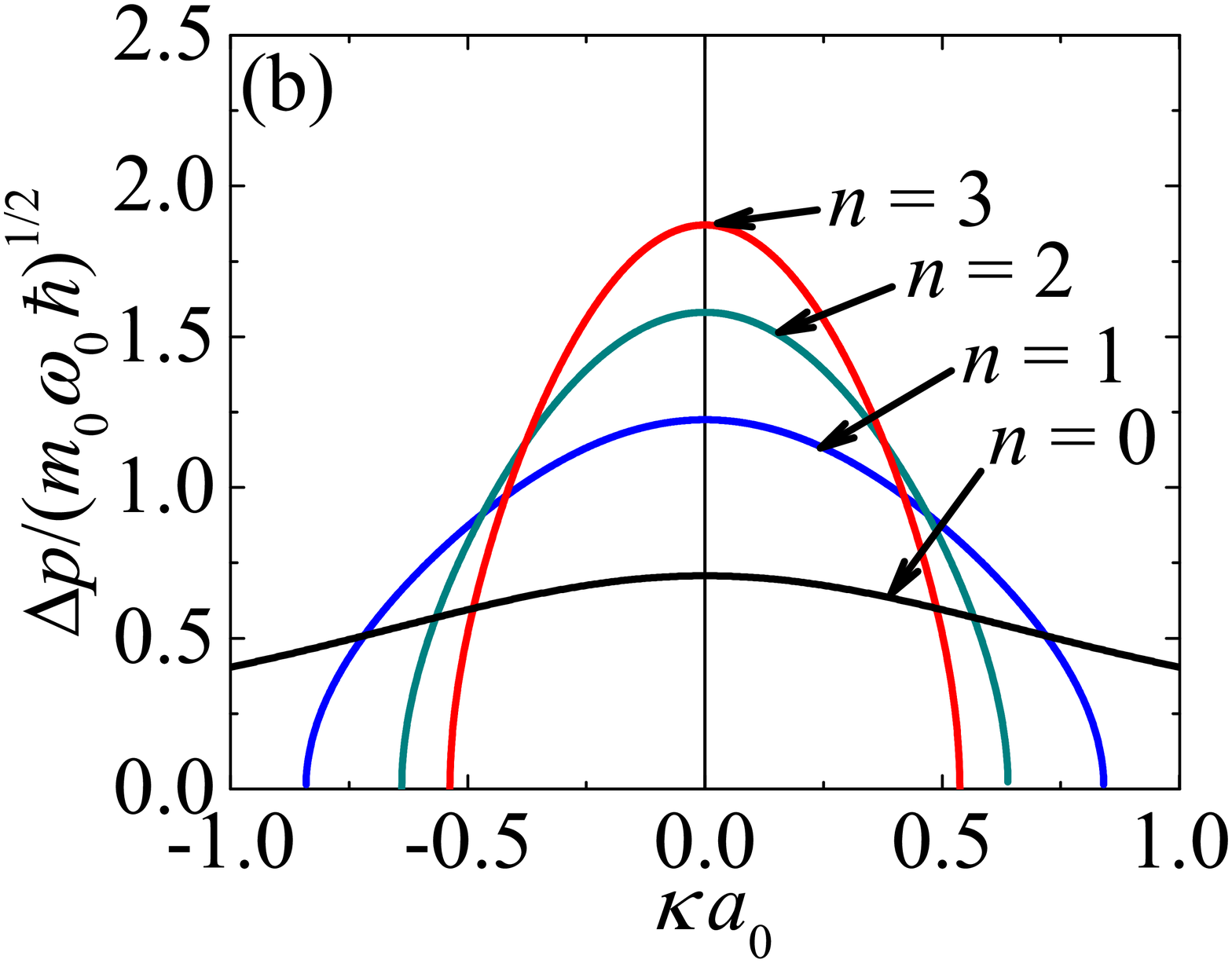}
\end{minipage}
\begin{minipage}[h]{0.32\linewidth}
\includegraphics[width=\linewidth]{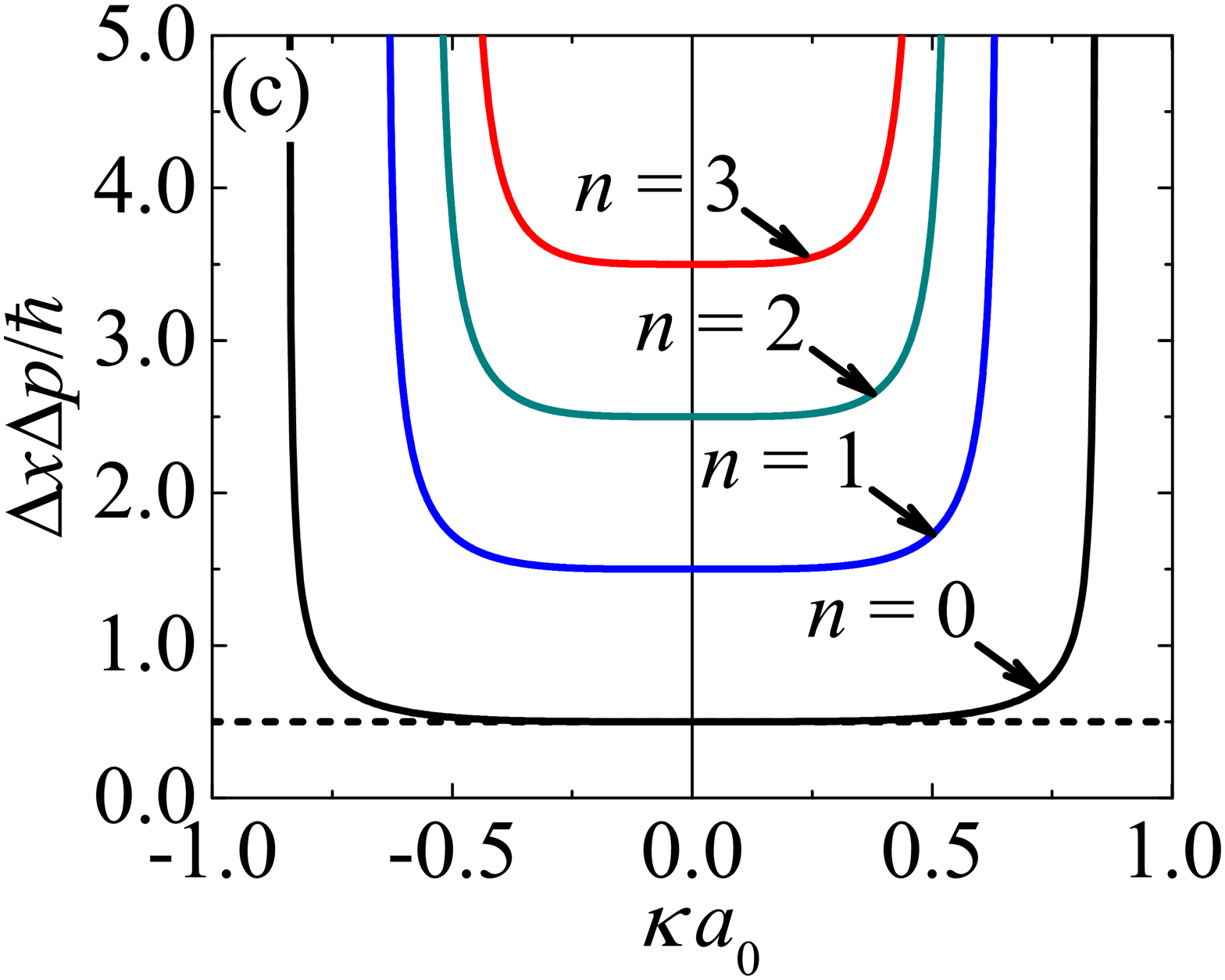}
\end{minipage}
\caption{\label{fig:9}
(Color online)
Uncertainty of (a) the position $\Delta x$,
(b) the momentum $\Delta p$,
and of the product (c) $\Delta x \Delta p$,
as function of $\kappa a_0$,
for the quantum states with $n=0$, $1$, $2$ and $3$.
The standard uncertainty relation
$\Delta x \Delta p = (n+\frac{1}{2})\hbar$
is recovered for $\kappa a_0 \rightarrow 0$.
}
\end{figure*}

In the limit $\kappa \rightarrow 0$, {\it i.e.} $z \rightarrow \infty$,
the usual cases
$\langle \hat{x}^2 \rangle
= \frac{\hbar}{m_0 \omega_0} \left( n+\frac{1}{2} \right)$
and
$\langle \hat{p}^2 \rangle
= m_0 \hbar \omega_0 \left( n+\frac{1}{2} \right)$
are recovered.
According to the principle of correspondence,
in the limit of large quantum numbers (or equivalently $\hbar \rightarrow 0$),
we have $E_n \rightarrow E$ and $\omega_\kappa \rightarrow \omega_0$,
and it is immediately verified that
Eqs.~(\ref{eq:x^2-quantum}) and (\ref{eq:p^2-quantum})
coincide with
Eqs.~(\ref{eq:x^2-med-osc}) and (\ref{eq:p^2-med-osc}), respectively.
Indeed, when $\hbar \rightarrow 0$ one obtains that
$z \approx 2m_0 \omega_0/\hbar \kappa^2 \gg 1$,
and we have
\begin{align}
\lim_{\hbar \rightarrow 0}\langle \hat{p}^2 \rangle &=
\lim_{\hbar \rightarrow 0}
m_0 E_n \left( 1 - \frac{2n+1}{z} \right)
\nonumber \\
&
=
\lim_{\hbar \rightarrow 0}
m_0 E_n \sqrt{1-\frac{2\kappa^2 E_n}{m_0 \omega_\kappa^2}
- \frac{\hbar^2 \kappa^4}{4m_0 \omega_\kappa^2}}
\nonumber \\
&=
m_0 E_n \sqrt{1-\frac{2\kappa^2 E_n}{m_0 \omega_\kappa^2}}
+\mathcal{O}({\hbar}^2).
\end{align}
The expectation values of the kinetic and potential energies satisfy
\begin{equation}
\label{eq:T-expected-value}
\langle \hat{T} \rangle
= E_n - \langle \hat{V} \rangle
= \frac{m_0 \omega_0^2}{2\kappa^2}
\frac{1}{\sqrt{1+\kappa^2 a_{n,\kappa}^2}}
\left(
\frac{\omega_0}{\omega_\kappa}-\frac{1}{\sqrt{1+\kappa^2 a_{n,\kappa}^2}}
\right),
\end{equation}
with
$E_n = \frac{m_0 \omega_0^2 a_{n,\kappa}^2}{
2(1+\kappa^2 a_{n,\kappa}^2)}$
and the quantum amplitude
\begin{equation}
a_{n,\kappa} =
	a_0 \left\{ \frac{
	\frac{\omega_\kappa}{\omega_0} (2n+1)
	-\kappa^2 a_0^2 \left( n^2 + n + \frac{1}{2} \right)}{
	1 - \kappa^2 a_0^2
	\left[
	\frac{\omega_\kappa}{\omega_0} (2n+1)
	-\kappa^2 a_0^2 \left( n^2 + n + \frac{1}{2} \right) \right]
	} \right\}^{1/2}.
\end{equation}
In the classical limit, one has that
$a_{n,\kappa} \rightarrow A_\kappa$ once $E_{n} \rightarrow E$, so
the expectation value (\ref{eq:T-expected-value})
recovers its classical average value
(\ref{eq:T-average-value}).
In Fig.~\ref{fig:9} we show the uncertainty relation of
the $\kappa$-deformed oscillator, along with the uncertainties
$\Delta x$ and $\Delta p$ of $x$ and $p$,
for the ground state and the first three excited ones.
As expected, while $\Delta x$ increases as
the dimensionless deformation parameter $\kappa a_0$
varies within the interval $[-1,1]$, $\Delta p$ decreases and viceversa.
In turn, this implies a generalized $\kappa$-uncertainty inequality
(Fig.~\ref{fig:9} (c)) which is an increasing function
of the quantum number $n$ and it also grows fast as $\kappa a_0$ varies.
The symmetry exhibited around the axis $\kappa a_0=0$
in the curves of the Fig.~\ref{fig:9} are
a consequence of the invariance of the mass function
(and then of the Hamiltonian too) given by Eq.~(\ref{eq:m(x)})
against the transformation $\kappa\rightarrow -\kappa$.

\section{Conclusions}
\label{sec:conclusions}

We have presented the
quantum and the classical mechanics
that results from assuming a position-dependent mass
related to the $\kappa$-algebra, which is
the mathematical background underlying $\kappa$-statistics.
Indeed, we have characterized both the quantum and classical
formalism of a particle
with a PDM determined univocally by the $\kappa$-algebra.
The consistency of the $\kappa$-deformed formalism is manifested in
the following arguments.

The $\kappa$-deformed Schr\"odinger equation
turns out to be equivalent to
a Schr\"odinger-like equation for a deformed wave-function provided
with a $\kappa$-deformed non-Hermitian momentum operator.
Within the $\kappa$-formalism
one can define deformed versions of
the continuity equation, the Fourier transform, etc.
In particular, a deformed Newton's second law in terms
of the deformed dual $\kappa$-derivative (Eq.
\eqref{eq:second_newton_law_generalized}) follows
in the classical limit.

We have illustrated the approach with
the problems of a particle confined in
an infinite potential well and
a $\kappa$-deformed oscillator
which is equivalent to the
Mathews-Lakshmanan oscillator
(in the standard space) or to the
P\"{o}sch-Teller potential problem
(in the $\kappa$-deformed space),
provided with the change of variable
$x\rightarrow x_\kappa$.
We have obtained the distributions for the classical
case as well as the eigenfunctions and eigenenergies
for the quantum case.
Although we have applied the mapping approach to a
$\kappa$-deformed space in order to
study the quantum Mathews-Lakshmanan oscillator,
it is important to mention that other equivalent
approaches can be found in the literature.
For instance, factorization methods, supersymmetry
and coherent states have also been
investigated for this nonlinear oscillator
(see \onlinecite{Amir-Iqbal-2016,Midya-Roy-2009,
Karthiga-2018,Amir-Iqbal-2016-CS,Tchoffo-2019}
and references therein).

Analogously to the quantum oscillator and
the Hermite polynomials,
the eigenvalues equation for the
$\kappa$-deformed quantum oscillator
is expressed in terms of the
Legendre polynomials.
Expectedly, in both examples we have reported the localization
and delocalization of the probability density functions
corresponding to the conjugated variables
$x$ and $p$, from which
the uncertainty relation follows
(Figs. \ref{fig:5} and \ref{fig:9}), with the particularity that
the lower bound is an increasing function of
the deformation parameter $\kappa$, satisfied by the ground state
and the first three excited ones.
This could be physically interpreted as if the quantum role
of the deformation (or equivalently, of the non-constant mass)
is to increase the intrinsic correlation between
the conjugated operators $\hat{x}$ and $\hat{p}$.
Also, for the case of the $\kappa$-deformed oscillator
we have studied the effect of the deformation parameter $\kappa$
on the phase space in the usual coordinates $(x,p)$
and in the deformed ones $(x_\kappa,\Pi_\kappa)$.
It is verified that for a certain range of values
of $\kappa$ the motion is unbounded (Fig.~\ref{fig:6}).

We consider that the techniques employed in this work could
stimulate the seek of other generalizations of classical
and quantum mechanical aspects,
as has been reported in recent researches
by means of the $q$-algebra.
\cite{Costa-Gomez-Santos-2020,
CostaFilho-Almeida-Farias-AndradeJr-2011,
Mazharimousavi-2012,
CostaFilho-Alencar-Skagerstam-AndradeJr-2013,
RegoMonteiro-Nobre-2013-PRA,
Costa-Borges-2014,
Arda-Server,
Braga-CostaFilho-2016,
Costa-Borges-2018,
Costa-Gomez-2020}

\section*{Acknowledgments}
ISG acknowledges support received from the National Institute of Science
and Technology for Complex Systems (INCT-SC), and from the
Conselho Nacional de Desenvolvimento Cient\'ifico e Tecnol\'ogico
(CNPq - Postdoctoral Fellowship 159799/2018-0)
(at Universidade Federal da Bahia), Brazil.
MP is grateful to CONICET and Universidad Nacional de La Plata, Argentina.

\section*{Data availability}
Data sharing is not applicable to this article as no new data were
created or analyzed in this study.



\end{document}